\documentclass[a4paper,11pt]{article}
\pdfoutput=1 

\usepackage{bm}

\usepackage{theorem}

\newtheorem{theorem}{Theorem}[section]

{\theorembodyfont{\normalfont}
 \newtheorem{definition}[theorem]{Definition}
 \newtheorem{example}{Example}[section]
 
 }

\newcommand{\qed}{\nobreak \ifvmode \relax \else
      \ifdim\lastskip<1.5em \hskip-\lastskip
      \hskip1.5em plus0em minus0.5em \fi \nobreak
      \vrule height0.75em width0.5em depth0.25em\fi}

\newcommand{\qeq}{\begin{equation}}
\newcommand{\eeq}{\end{equation}}
\newcommand{\bea}{\begin{eqnarray*}}
\newcommand{\eea}{\end{eqnarray*}}
\newcommand{\beqa}{\begin{eqnarray}}
\newcommand{\eeqa}{\end{eqnarray}}

\newcommand{\bX}{\boldsymbol{X}}

\def\ba{{\bm{A}}}

\def\bw{{\bm{W}}}

\def\by{{\bm{Y}}}
\def\bz{{\bm{Z}}}
\def\bq{{\bm{Q}}}
\def\bp{{\bm{P}}}

\def\bR{{\mathbb{R}}}

\def\bZ{{\mathbb{Z}}}

\newcommand{\calD}{{\mathcal D}}

\newcommand{\calF}{{\mathcal F}}

\newcommand{\calL}{{\mathcal L}}
\newcommand{\calM}{{\mathcal M}}

\newcommand{\calO}{{\mathcal O}}

\newcommand{\calR}{{\mathcal R}}
\newcommand{\calS}{{\mathcal S}}

\newcommand{\Map}{{\rm Map}}

\newcommand{\sbv}[2]{{\{{{#1},{#2}}\}}}
\newcommand{\courant}[2]{{[{{#1},{#2}}]_D}}
\newcommand{\courante}[2]{{[{{#1},{#2}}]_E}}

\newcommand{\courantr}[2]{{[{{#1},{#2}}]_D^{\pi}}}

\newcommand{\rhoe}{{\rho_{(E)}}}
\newcommand{\rhoa}{{\rho_{(A)}}}

\newcommand{\rhott}{\rho_{T \oplus T^*}}

\newcommand{\bracket}[2]{\langle #1,\,#2\rangle}

\newcommand{\inner}[2]{{({{#1},{#2}})}}

\newcommand{\rd}{\mathrm{d}}

\newcommand{\gomega}{{\omega}_{grad}}
\newcommand{\ggomega}{{\omega_{AKSZ}}}

\newcommand{\Thetac}{{\Theta_C}}
\newcommand{\Thetas}{{\Theta_S}}
\newcommand{\Thetaa}{{\Theta_A}}
\newcommand{\Thetae}{{\Theta_E}}
\newcommand{\Thetaf}{{\Theta_F}}

\newcommand{\baS}{{{}^A S}}
\newcommand{\beS}{{{}^E S}}

\newcommand{\enablab}{{{}^E \nabla^{bas}}}
\newcommand{\anablab}{{{}^A \nabla^{bas}}}
\newcommand{\banabla}{{}^A \stackrel{\bullet}{\nabla}}

\usepackage{jheppub} 

\usepackage[T1]{fontenc} 

\title{\boldmath Gauged Courant sigma models}

\author[a,b]{Noriaki Ikeda}

\affiliation[a]{Research Institute for Mathematical Sciences, 
Kyoto University \\
Kyoto 606-8502, Japan}
\affiliation[b]{Department of Mathematical Sciences,
Ritsumeikan University \\
Kusatsu, Shiga 525-8577, Japan}

\emailAdd{nikeda@se.ritsumei.ac.jp}

\abstract{We propose a new class of sigma models based on Courant sigma models. We refer to these models as gauged Courant sigma models (GCSMs). By introducing additional gauge symmetries, such as those associated with a Lie group, a Lie groupoid (or Lie algebroid), and a Courant algebroid on the target space, Courant sigma models are extended to gauged sigma models of AKSZ type. The consistency of the theory is ensured by identities among geometric quantities on Lie algebroids and Courant algebroids, such as curvatures and torsions, which can be interpreted as flatness conditions on the target space. We also analyze geometric structures of GCSMs in the presence of fluxes and boundaries.}

\begin{document} 
\maketitle
\flushbottom

\section{Introduction}
\noindent
A Courant sigma model is a topological sigma model defined on 
a three-dimensional manifold \cite{Ikeda:2002wh, Hofman:2002rv, 
Roytenberg:2006qz}, 
and it can be regarded as a generalization of 
Chern-Simons gauge theory.
The target space of the model is a Courant algebroid $E$ \cite{LWX}
over a smooth manifold $M$.
This model is also an example of AKSZ sigma models 
\cite{Alexandrov:1995kv} constructed 
on mapping spaces between two graded manifolds
and there are many mathematical and physical applications.

In this paper, we consider \textit{gauging} of the Courant sigma model.
Gauging a sigma model means promoting a global symmetry to a local 
gauge symmetry by introducing the corresponding gauge fields.
From a mathematical viewpoint, this corresponds to considering a target space equipped 
with group or groupoid actions in addition to the original Courant algebroid structure.
We call the resulting model obtained by gauging the Courant sigma model
a \textit{gauged Courant sigma model} (GCSM).

In our previous work \cite{Ikeda:2023pdr}, we analyzed the gauging of 
the Poisson sigma model \cite{Ikeda:1993fh, Schaller:1994es} 
and the Dirac sigma model \cite{Kotov:2004wz} by Lie groupoids (Lie algebroids)
including Lie groups (Lie algebras).
Target space geometry corresponds to geometry of Poisson manifolds 
with a group or groupoid actions.
In that work, we formulated sigma model descriptions of Hamiltonian $G$-spaces
and their generalizations, namely Hamiltonian Lie algebroid structures 
\cite{Blohmann:2018}.
These models are two-dimensional AKSZ sigma models, that is, two-dimensional analogues of the Courant sigma model.
They are referred to as the gauged Poisson sigma model (GPSM) and the gauged Dirac sigma model (GDSM).

The compatibility conditions between gauge symmetries and Poisson or Dirac structures are described by flatness conditions on curvature-type objects, such as the basic curvature of Lie algebroid connections.
Moreover, when considering the GPSM and GDSM on manifolds with boundaries, 
boundary conditions are determined by Hamiltonian Lie algebroids \cite{Blohmann:2018} over Poisson manifolds \cite{Blohmann:2023} and Dirac structures \cite{Ikeda:2023pdr}.
Another approach to gauging the Poisson sigma model has been proposed in \cite{Zucchini:2008cg}.

Since the Poisson sigma model is an example of an AKSZ sigma model, it is natural to consider the gauging of more general AKSZ sigma models.
The Poisson sigma model is based on a QP-manifold of degree one.
In this paper, we study the gauging of the Courant sigma model, which is an AKSZ sigma model based on a QP-manifold of degree two.
For graded geometry including QP-manifolds, refer to \cite{Roytenberg:2006qz, Cattaneo:2010re, Ikeda:2012pv, Cueca:2025aej} and references therein.

Group-like objects of Courant algebroids are Lie $2$-groupoids.
One should therefore consider actions of Lie 
$2$-groupoids, which serve as the appropriate group-like objects.
See \cite{LiBlandSevera, MehtaTang1, MehtaTang2} for the integration of 
Courant algebroids to Lie $2$-groupoids.
Another proposal for group-like objects associated with Courant algebroids is given by 
Lie rackoids \cite{LaurentGengouxWagemann}.
In this paper, however, we restrict ourselves to infinitesimal data, 
namely Courant algebroids, since the Courant algebroid structure--or equivalently, 
a QP-manifold of degree two--is sufficient to construct an AKSZ action functional.
The analysis of global group-like structures in the GCSM is left for future work.

Another important problem is the quantization. The quantization of the GCSM 
is also left for future work.

A Courant sigma model is constructed as an AKSZ sigma model, that is, it is based on a 
QP-manifold or, equivalently, a differential graded (dg) symplectic manifold.
This construction relies on the fact that a Courant algebroid $E$ 
is in one-to-one correspondence with a QP-manifold of degree two,
$T^*[2]E[1]$ \cite{Roy01}.
By the AKSZ construction, a QP-manifold structure on the mapping space
$\Map(T[1]\Sigma, T^*[2]E[1])$
is induced from the QP-structure on $T^*[2]E[1]$.
The resulting theory is equivalent to the Batalin-Vilkovisky formulation 
\cite{BV1, BV2} of the Courant sigma model.
We employ the same construction for the gauged Courant sigma model in this paper.

To ensure the homological condition $Q^2=0$, where 
$Q$ is a degree-one vector field on the mapping space, 
certain geometric conditions corresponding to flatness of curvature-type 
objects are required.
If such flatness conditions are not imposed, 
the theory should instead be regarded as an equivariant version of 
an AKSZ theory, in which 
$Q^2=0$ holds only up to groupoid actions.
Such generalizations of AKSZ theories and equivariant versions of BV and AKSZ theories 
have been discussed in 
\cite{Bonechi:2012kh, Bonechi:2019dqk, Grigoriev:2024ncm}.
Gauged Courant sigma models provide concrete examples of this framework.

Moreover, in Section \ref{sec:CSMflux}, and
Apendix \ref{sec:SCSMCAflux}, \ref{sec:GCSMLAflux} and 
\ref{sec:GCSMCAflux},
we consider deformations of gauged Courant sigma models
by fluxes and boundary terms.
By introducing various fluxes, consistency conditions are deformed.
Conditions are regarded as generalizations of conditions of fluxes 
in string theory \cite{Grana:2008yw, Heller:2016abk}.

If we consider boundary terms, the homological condition $Q^2=0$
imposes conditions of generalizations of momentum maps on boundary terms,
including homotopy moment maps \cite{Callies:2013jbu}
and homotopy momentum sections \cite{Hirota:2021isx} as special cases.
As a result, we obtain generalizations of momentum maps to Courant 
algebroid settings.

This paper is organized as follows.
In Section 2, we summarize Courant sigma models and the AKSZ formulation.
In Section 3, we construct various versions of gauged Courant sigma models.
Various types of Courant algebroids, the standard Courant algebroid and 
general Courant algebroids, and various type of gauging, Lie algebroids and 
Courant algebroids are analyzed.
In Section 4, we generalize models by introducing fluxes and boundaries.
In the Appendix, we summarize definitions, formulas and 
local coordinate expressions of Lie algebroids, Courant algebroids 
and Q(P)-manifolds. 
Moreover calculations of models with fluxes and with boundaries are summarized 
in the Appendix D-F.

\paragraph{Notation}

$i, j, k$, etc are for indices of local coordinates in a target base manifold 
$M$, and its tangent and cotangent bundle.
$A, B, C$, etc are for indices of local coordinates of the fiber of a vector bundle $E$, which is a whole Courant algebroid.
$a, b, c$, etc are for indices of local coordinates of 
the fiber of a gauging vector bundle $A$, which gives gauging symmetries.
$k_{AB}$ is a local coordinate expression of a fiber metric of a vector bundle $E$.
$m_{ab}$ is a local coordinate expression of a fiber metric of a vector bundle $A$.

\section{Courant Sigma models}\label{sec:CSM}
In this section, we summarize Courant sigma models including explanations of our notation.
One can refer to basic facts of Courant algebroids and Q-manifolds in 
Section \ref{app:courantalgebroid} and Section \ref{app:qmanifold}
in Appendix.

Let $\Sigma$ be a three-dimensional manifold and
$E$ be a vector bundle  over $M$ with a fiber metric, $\bracket{-}{-}$.
We consider a map from $\Sigma$ to $M$, $X: \Sigma \rightarrow M$.
Moreover, we introduce a $1$-form on $\Sigma$ taking a value on the pullback 
of $E$, $Y \in \Omega^1(\Sigma, X^*E)$
and a $2$-form on $\Sigma$ taking a value on the pullback 
of $T^*M$, $Z \in \Omega^2(\Sigma, X^*T^*M)$.

The action functional of a Courant sigma model \cite{Ikeda:2002wh, Roytenberg:2006qz} is 
\begin{eqnarray}
S_C &=& \int_{\Sigma} 
\left[- \inner{Z}{\rd X} + \frac{1}{2} \bracket{Y}{\rd Y}
+ \inner{(\rho \circ X) (Y)}{Z} + \frac{1}{3!} X^*H(Y, Y, Y)\right],
\label{CSMaction}
\end{eqnarray}
where $\rd$ is the de Rham differential on $\Sigma$,
$\inner{-}{-}$ is the pairing of $TM$ with $T^*M$,
$\rho:E \rightarrow TM$ is a bundle map and 
$H$  is a tensor field $H \in \Gamma(\wedge^3 E^*)$.

If we introduce local coordinates on the target vector bundle $E$,
the action functional is given by
\begin{align}
S_C &= \int_{\Sigma} 
\left[- Z_i \wedge \rd X^i + \frac{1}{2} k_{AB} Y^A \wedge \rd Y^B
+ \rho^i_A(X) Z_i \wedge Y^A
+ \frac{1}{3!} H_{ABC}(X) Y^A \wedge Y^B \wedge Y^C \right],
\label{CSMaction2}
\end{align}
where $i = 1, \ldots, \dim(M)$ is an index of the base manifold $M$,
the tangent bundle $TM$ and  the cotangent bundle $T^*M$.
$A, B, C = 1, \ldots, \mathrm{rank}(E)$ are indices of the fiber of the vector bundle $E$.
For a basis $e_A$ of the fiber of $E$,
$k_{AB} = \bracket{e_A}{e_B}$  is a local coordinate expression of the fiber metric
$\bracket{-}{-}$, $\rho_A^i(x) \partial_i = \rho(e_A)$ is the bundle map
and 
$
H_{ABC}(x) = H(e_A, e_B, e_C) $.

The action functional $S_C$ is consistent if and only if 
$k_{AB}, \rho_A^i, H_{ABC}$ satisfy the axiom of a Courant algebroid, 
thus the model is called a Courant sigma model.

The Courant sigma model is an AKSZ sigma model. 
There exists a construction based on differential graded (dg) symplectic 
manifolds called QP-manifolds.
It is equivalent to Batalin-Vikovisky (BV) formalism of gauge theories 
in physics.

We start at a QP-manifold structure on the graded manifold $T^*[2]E[1]$
with a graded symplectic form of degree two $\gomega$ and homological function 
$\Thetac$. By definition, $\Thetac$ satisfies the homological condition,
$\sbv{\Thetac}{\Thetac}=0$ by a graded Poisson bracket $\sbv{-}{-}$ of degree 
$-2$ induced from $\gomega$.

In the case of Courant algebroids,
the graded symplectic form is
\begin{align}
\gomega &= \delta x^i \wedge \delta z_i 
+ \frac{1}{2} \delta \eta^A \wedge \delta (k_{AB} \eta^B),
\label{gsymplectic1}
\end{align}
and the homological function of degree three is
\begin{align}
\Thetac &= \rho^i_A(x) z_i \eta^A 
+ \frac{1}{3!} H_{ABC}(x) \eta^A \eta^B \eta^C,
\label{CAhomologicalfn1}
\end{align}
where $(x^i, z_i, \eta^A)$ are local coordinates 
on $T^*[2]E[1]$ of degree $(0, 2, 1)$,
and $\delta$ is the graded de Rham differential on the graded manifold.
From the definition of the QP-manifold,
the homological function satisfies the condition,
\begin{align}
\sbv{\Thetac}{\Thetac} &= 0,
\label{homological1}
\end{align}
If and only if $\Thetac$ satisfies Eq.~\eqref{homological1},
the vector bundle $E$ has a Courant algebroid structure.
Concrete correspondence appears in Example \ref{CAQP}.

The space of fields in the Courant sigma model is the mapping space from 
$T[1]\Sigma$ to $T^*[2]E[1]$, $\Map(T[1]\Sigma, T^*[2]E[1])$.
Local coordinates on $\Map(T[1]\Sigma, T^*[2]E[1])$ are called superfields. 
We denotes superfields by bold letters 
$(\bX^i(\sigma, \theta), \bz_i(\sigma, \theta), \by^A(\sigma, \theta))$
of total degree $(0, 2, 1)$, which correspond to local coordinates 
$(x^i, z_i, \eta^A)$ on $T^*[2]E[1]$.
Let $(\sigma^{\mu}, \theta^{\mu})$ be local coordinates on $T[1]\Sigma$ 
of degree $(0, 1)$. 
The theory has bidegree, degree in $T[1]\Sigma$ and degree in $T^*[2]E[1]$.
Superfields are expanded by $\theta^{\mu}$ as
\begin{align}
\bX^i(\sigma, \theta) &= X^{(0)i}(\sigma)
+ \theta^{\mu} X^{(1)i}_{\mu}(\sigma) 
+ \frac{1}{2} \theta^{\mu} \theta^{\nu} X^{(2)i}_{\mu\nu}(\sigma) 
+ \frac{1}{3!} \theta^{\mu} \theta^{\nu} \theta^{\lambda} 
X^{(3)i}_{\mu\nu\lambda}(\sigma),
\\
\bz_i(\sigma, \theta) &= Z_i^{(0)}(\sigma)
+ \theta^{\mu} Z^{(1)}_{\mu i}(\sigma) 
+ \frac{1}{2} \theta^{\mu} \theta^{\nu} Z^{(2)}_{\mu\nu i}(\sigma) 
+ \frac{1}{3!} \theta^{\mu} \theta^{\nu} \theta^{\lambda} 
Z^{(3)}_{\mu\nu\lambda i}(\sigma),
\\
\by^A(\sigma, \theta)) &=Y^{(0)A}(\sigma)
+ \theta^{\mu} Y^{(1)A}_{\mu}(\sigma) 
+ \frac{1}{2} \theta^{\mu} \theta^{\nu} Y^{(2)A}_{\mu\nu}(\sigma) 
+ \frac{1}{3!} \theta^{\mu} \theta^{\nu} \theta^{\lambda} 
Y^{(3)A}_{\mu\nu\lambda}(\sigma).
\end{align}
Degree zero coefficient functions of expansions 
correspond to classical fields, i.e.,
$X^i = X^{(0)i}(\sigma)$, $Z_{\mu\nu i} = Z^{(2)}_{\mu\nu i}(\sigma)$
and $Y^A_{\mu} = Y^{(1)A}_{\mu}(\sigma)$ are classical fields.
Degree nonzero parts are ghosts and antifields.
The BV action functional is completely constructed by these superfields,
which is called the AKSZ formulation.
The BV action functional is
\begin{align}
S_C &= \int_{T[1]\Sigma} \rd^3 \sigma \rd^3 \theta
\left[- \bz_i \rd \bX^i + \frac{1}{2} k_{AB} \by^A \rd \by^B
+ \rho^i_A(\bX) \bz_i \by^A
+ \frac{1}{3!} H_{ABC}(\bX) \by^A \by^B \by^C \right],
\label{CSMAKSZaciton}
\end{align}
where $\rd$ is replaced to the super derivative on $T[1]\Sigma$,
$\rd = \theta^{\mu} \tfrac{\partial}{\partial \sigma^{\mu}}$.
Throughout this paper, we use this formalism, thus, 
$\rd$ on $T[1]\Sigma$ is always the super derivative in later sections.

As a special case, we start at the QP-manifold induced from 
a standard Courant algebroid $TM \oplus T^*M$
in Example \ref{standardCA}.
From the graded symplectic form \eqref{gsymplectic2} and 
the homological function \eqref{CAhomologicalfn2} 
in Example \ref{standardCAQP},
the AKSZ-BV action functional is obtained as
\begin{align}
S_C &= \int_{T[1]\Sigma} \rd^3 \sigma \rd^3 \theta
\left[- \bz_i \rd \bX^i + \bp_i \rd \bq^i
+ \bz_i \bq^i
+ \frac{1}{3!} H_{ijk}(\bX) \bq^i \bq^j \bq^k \right],
\label{SCSMAKSZaciton}
\end{align}
where $\bq^i(\sigma, \theta)$ and 
$\bp_i(\sigma, \theta)$ are degree one superfields corresponding to
coordinates of degree one $(q^i, p_i)$ on the target space $T^*[2]T^*[1]M$.
Eq.~\eqref{SCSMAKSZaciton} is called the standard Courant sigma model (SCSM).

\section{Gauged Courant sigma models}
In this section, we consider concrete gauging of Courant sigma models.
We consider several kinds of gauging of Courant sigma models,
standard Courant sigma models (SCSMs) 
and general Courant sigma models (GCSMs)
with gauging by Lie algebroids and Courant algebroids.

\subsection{Standard Courant sigma models with Lie algebroid gauging}
\label{sec:SCSMLA}
Suppose that the target space is a standard Courant algebroid $TM \oplus T^*M$.
The QP-manifold is $T^*[2]T^*[1]M$.
Then the AKSZ action is the standard Courant sigma models (SCSM)
\eqref{SCSMAKSZaciton}.

We consider gauging by Lie algebroids.
Note that ordinary local gauge symmetries of Lie groups are special cases of 
gauging by Lie algebroids. 
Because infinitesimal gauge symmetries of Lie group are
Lie algebra gauge symmetries, which always give action 
Lie algebroid structures.

Let $A$ be a Lie algebroid over $M$. For gauging of the SCSM, 
$T^*[2]T^*[1]M$ is extended to $T^*[2](T^*[1]M \oplus A[1])$.
First, we consider a QP-manifold structure on $T^*[2](T^*[1]M \oplus A[1])$.
Next, the AKSZ construction gives a gauged Courant sigma model from 
the QP-manifold structure.

We take local coordinates on $T^*[2]A[1]$, $(a^a, w_a)$ of degree $(1,1)$
with $a = 1, \ldots, \mathrm{rank}(A)$,
where $a^a$ is a local coordinate of the fiber of $A[1]$ and 
$w_a$ is a canonical conjugate local coordinate on $T^*[2]$.
Since $T^*[2](T^*[1]M \oplus A[1])$ is a graded cotangent bundle, 
there exists the canonical graded symplectic form of degree two as,
\begin{align}
\gomega &= 
\delta x^i \wedge \delta z_i 
+ \delta q^i \wedge \delta p_i
+ \delta a^a \wedge \delta w_a
\label{gsymplectic3}
\end{align}
From Eq.~\eqref{gsymplectic3}, we obtain graded Poisson brackets for canonical quantities,
\begin{align}
\sbv{x^i}{z_j} &= \delta^i_j,
\\
\sbv{q^i}{p_j} &= \delta^i_j,
\\
\sbv{a^a}{w_b} &= \delta^a_b,
\end{align}
We introduce covariant local coordinates to calculate gauging 
by the Lie algebroid $A$.
In order to make covariant QP-structures for the Lie algebroid action $A$,
an ordinary connection $\nabla$ on $A$ is introduced. 
The corresponding connection $1$-form is 
$\omega = \omega^a_{bi} \rd x^i \otimes e_a \otimes e_b$ 
such that $\nabla e_a = - \omega_{ai}^b \rd x^i \otimes e_b$ for 
basis $e_a$ in $\Gamma(A)$\footnote{For complete covariant expressions of the theory,
we have to introduce an affine connection $\Gamma_{ij}^k$ on $TM$
for the standard Courant algebroid $TM \oplus T^*M$.
However they are not introduced in this paper since the total QP-structure, 
the graded symplectic form and the homological function, do not depend on 
the affine connection. As the result, gauging conditions 
does not depend on an affine connection on $TM$,
and only depend on a connection on $A$.
}.
If a Lie algebroid $A$ is an action Lie algebroid $A = M \times \mathfrak{g}$,
i.e., symmetries induced from a Lie algebra $\mathfrak{g}$,
we do not need a connection since gauging does not depend on the connection.
One can refer to Appendix \eqref{connectionLA} for curvatures, $A$-connections, $A$-torsions, basic curvatures, etc. in Lie algebroids.

After gauging of the Lie algebroid $A$, the resulting theory is invariant under 
local coordinate transformations on $T^*[2]A[1]$.
As explained in Section \ref{manicovariant} in Appendix,
$z_i$ is not transformed covariantly under local coordinate transformations
on $T^*[2]A[1]$.
We can change $z_i$ to the covariantized local coordinate 
$z_i^{\nabla} := z_i + \omega_{ai}^b a^a w_b$.
In fact, from Eqs.~\eqref{coordinatetransf1}--\eqref{coordinatetransf3},
we can check that $z_i^{\nabla}$ is covariant, where $\eta^A = (a^a, w_a)$.
The Liouville $1$-form $\vartheta$ for the graded symplectic form 
such that $\gomega = - \delta \vartheta$ is
\begin{align}
\vartheta &= 
z_i \delta x^i 
- p_i \delta q^i 
- w_a \delta a^a 
\nonumber \\
& = z_i^{\nabla} \delta x^i 
- p_i \delta q^i 
- w_a D a^a,
\label{Liouville3}
\end{align}
where $D a^a = \delta a^a - \omega^a_{bi} a^b \delta x^i$.
In fact, we can prove $\vartheta$ is invariant under local coordinate 
transformations on the graded manifold $T^*[2](T^*[1]M \oplus A[1])$,
and as a result $\gomega$ is invariant.

Poisson brackets between $z_i^{\nabla}$ and other coordinates are
\begin{align}
\sbv{z_i^{\nabla}}{a^a} &= \omega_{bi}^a a^b,
\\
\sbv{z_i^{\nabla}}{w_a} &= - \omega_{ai}^b w_b,
\\
\sbv{z_i^{\nabla}}{z_j^{\nabla}} &= R_{ijb}^a w_a a^b,
\end{align}
and other Poisson brackets are zero,
where $R$ is a curvature in Eq.~\eqref{curv} or Eq.~\eqref{curvatureona}.

Homological functions in the QP-manifold for a standard Courant algebroid 
and a Lie algebroid are in Eqs.~\eqref{CAhomologicalfnl}
and \eqref{CAhomologicalfn2},
\begin{align}
\Thetas &= z_i q^i + \frac{1}{3!} H_{ijk}(x) q^i q^j q^k,
\\
\Thetaa &= \rho^i_a(x) z_i a^a + 
\frac{1}{2} C_{ab}^c(x) a^a a^b w_c.
\end{align}
However they are not invariant under local 
coordinate transformations on $T^*[2]A[1]$.

Covariantized homological functions, which are invariant,
are ones replaced by covariant coordinates $z_i^{\nabla}$ 
and covariant tensors as
\begin{align}
\Thetas^{\nabla} &= \Thetas
+ \omega_{bi}^c a^b w_c q^i 
= z_i^{\nabla} q^i 
+ \frac{1}{3!} H_{ijk}(x) q^i q^j q^k,
\label{thetas01}
\\
\Thetaa^{\nabla} &= \Theta_G = \rho^i_a(x) z_i^{\nabla} a^a 
- \frac{1}{2} T_{ab}^c(x) a^a a^b w_c.
\label{thetaa01}
\end{align}
The total function is
$\Theta^{\nabla} := \Thetas^{\nabla} + \Thetaa^{\nabla}$,
where $T_{ab}^c$ is the $A$-torsion on the Lie algebroid $A$.
Note that $\Thetaa^{\nabla} = \Thetaa$ but 
$\Thetas^{\nabla} \neq \Thetas$, thus 
homological conditions including $\Thetas^{\nabla}$ are not 
satisfied any more.
In fact, the graded Poisson bracket
$\sbv{\Theta^{\nabla}}{\Theta^{\nabla}}$ are as follows.
\begin{align}
\sbv{\Thetas^{\nabla}}{\Thetas^{\nabla}} 
& = - R_{ijc}^d q^i q^j a^c w_d,
\label{cme01} \\
\sbv{\Thetaa^{\nabla}}{\Thetaa^{\nabla}} &= 0,
\label{cme02} \\
\sbv{\Thetas^{\nabla}}{\Thetaa^{\nabla}} 
&= \nabla_i \rho^j_a z_j^{\nabla} a^a q^i
- \frac{1}{2} S_{iab}^c a^a a^b w_c q^i
- \frac{1}{3!} \rho^l_{a} \partial_{l} H_{ijk} a^a q^i q^j q^k.
\label{cme03}
\end{align}
where local coordinate expressions of the curvature \eqref{curvatureona}, 
the $A$-torsion \eqref{atorsionona}, 
and the basic curvature \eqref{bcurvatureona} are used.
The total function is $\Theta^{\nabla} = \Thetas^{\nabla} + \Thetaa$.
Therefore $\sbv{\Theta^{\nabla}}{\Theta^{\nabla}}=0$ is satisfied
if the following conditions hold,
\begin{align}
& R_{ija}^b =0,
\label{homological01}
\\
& S_{iab}^c =0,
\label{homological02}
\\
& \nabla_i \rho^j_a =0,
\label{homological03}
\\
& \rho^l_{a} \partial_{l} H_{ijk} = 0.
\label{homological04}
\end{align}
By combining Eqs.~\eqref{homological03} and \eqref{homological04},
we obtain the equation ${}^A \rd H =0$.
Note that ${}^A \rd H = {}^A \nabla H$ under $\nabla \rho =0$.
The we summarize the result as follows.
\begin{theorem}
$\Theta^{\nabla}$ is homological $\sbv{\Theta^{\nabla}}{\Theta^{\nabla}}=0$ 
if and only if
the curvature and the basic curvature vanish $R = {}^A S =0$,
the anchor map is horizontal $\nabla \rho = 0$,
and $H$ satisfies ${}^A \nabla H =0$.
\end{theorem}

We can consider the structure is a \textit{equivariant} QP-structure.
A homological vector field $Q$ is extended to an equivariant 
vector field such that 
$Q^{\nabla} = Q + \iota_{\Omega} = \sbv{\Theta^{\nabla}}{-}$
with a 'connection' $\Omega$, where $Q^{\nabla} = \sbv{\Theta^{\nabla}}{-}$.
$(Q^{\nabla})^2 = \calL_{\calF}$ with a 'curvature' $\calF$.
Here $\calF$ is the vector field induced from $(Q^{\nabla})^2$.
The $Q^{\nabla}$-complex gives an equivariant cohomology up to Lie algebroid action.
Equivariant versions of BV formalisms and AKSZ theories are analyzed in 
\cite{Bonechi:2019dqk}.
Physical theories with $Q^2 \neq 0$ appear in double field theories
\cite{Chatzistavrakidis:2018ztm, Carow-Watamura:2018iau}.

From the above data, an AKSZ action $S_S^{\nabla}$ is constructed.
The kinetic term is induced from the Liouville $1$-form \eqref{Liouville3} as
\begin{align}
& \int_{T[1]\Sigma} \rd^3 \sigma \rd^3 \theta
\left[- \bz_i \rd \bX^i 
- \bp_i \rd \bq^i
- \bw_a \rd \ba^a \right]
\nonumber \\ &=
\int_{T[1]\Sigma} \rd^3 \sigma \rd^3 \theta
\left[- \bz_i^{\nabla} \rd \bX^i 
- \bp_i \rd \bq^i
- \bw_a D \ba^a \right],
\end{align}
where $D \ba^a = \rd \ba^a - \omega^a_{bi} \ba^b \rd \bX^i$.
Other terms are induced from homological functions
\eqref{thetas01} and \eqref{thetaa01}.
As a covariant modification of Eq.~\eqref{SCSMAKSZaciton},
the covariantized deformations of the standard Courant sigma model is given by
\begin{align}
S_S^{\nabla} &= \int_{T[1]\Sigma} \rd^3 \sigma \rd^3 \theta
\left[- \bz_i^{\nabla} \rd \bX^i 
+ \bp_i \rd \bq^i
+ \bz_i^{\nabla} \bq^i
+ \frac{1}{3!} H_{ijk}(\bX) \bq^i \bq^j \bq^k \right],
\end{align}
We add gauging terms to action $S_S^{\nabla}$ induced from 
the Liouville $1$-form and $\Thetaa$,
\begin{align}
S_A &= \int_{T[1]\Sigma} \rd^3 \sigma \rd^3 \theta
\left[\bw_a D \ba^a
+ \rho^i_a(\bX) \bz_i \ba^a 
+ \frac{1}{2} C_{ab}^c(\bX) \ba^a \ba^b \bw_c \right]
\nonumber \\
&= \int_{T[1]\Sigma} \rd^3 \sigma \rd^3 \theta
\left[\bw_a D \ba^a
+ \rho^i_a(\bX) \bz_i^{\nabla} \ba^a 
- \frac{1}{2} T_{ab}^c(\bX) \ba^a \ba^b \bw_c \right].
\end{align}
The resulting AKSZ action of a gauged Courant Sigma model 
of the standard Courant algebroid with a Lie algebroid gauging is
\begin{align}
S &= S_S^{\nabla} + S_A 
\nonumber \\ 
&=\int_{T[1]\Sigma} \rd^3 \sigma \rd^3 \theta
\left[- \bz_i ^{\nabla} \rd \bX^i 
+ \bp_i \rd \bq^i 
+ \bw_a D \ba^a
+ \bz_i^{\nabla} \bq^i
+ \rho^i_a(\bX) \bz_i^{\nabla} \ba^a 
\right.
\nonumber \\
& 
\left.
+ \frac{1}{3!} H_{ijk}(\bX) \bq^i \bq^j \bq^k 
- \frac{1}{2} T_{ab}^c(\bX) \ba^a \ba^b \bw_c \right],
\label{actionSCSMLA}
\end{align}
If gauging symmetry is a Lie group, the infinitesimal symmetry is 
a Lie algebra. Let $\mathfrak{g}$ be the symmetry Lie algebra.
Then $C_{ab}^c$ is globally constant,i.e., the structure constant of 
$\mathfrak{g}$, and a vector field $\rho_a := \rho^i_a(x)\partial_i$ 
is the infinitesimal action of $\mathfrak{g}$ as a differential operator,
so called the Killing vector field.
Then, $A = M \times \mathfrak{g}$ has an action Lie algebroid structure.

\subsection{Standard Courant sigma models with Courant algebroid gauging}
\label{sec:SCSMCA}
In this section, we consider gauging of the standard Courant sigma model 
by another Courant algebroid $E$ on the same base manifold $M$.
For gauging, we extent the graded manifold to
$T^*[2](T^*[1]M \oplus E[1])$, where $E$ is a general Courant algebroid.
Let $\bracket{e_a}{e_b} = m_{ab}$ 
be the local coordinate expression of the inner product 
of the Courant algebroid for the basis of the fiber $e_a \in \Gamma(E)$.
Take a degree one local coordinate on the fiber of $E[1]$ by 
$a^a$, $a= 1, \ldots, \mathrm{rank}(E)$.
Then, the canonical graded symplectic form of degree two on the graded cotangent bundle 
$T^*[2](T^*[1]M \oplus E[1])$ is
\begin{align}
\gomega &= 
\delta x^i \wedge \delta z_i 
+ \delta q^i \wedge \delta p_i
+ \frac{1}{2} \delta a^a \wedge \delta (m_{ab} a^b).
\label{gsymplectic4}
\end{align}
From the symplectic form \eqref{gsymplectic4}, we obtain canonical
graded Poisson brackets,
\begin{align}
\sbv{x^i}{z_j} &= \delta^i_j,
\\
\sbv{q^i}{p_j} &= \delta^i_j,
\\
\sbv{a^a}{a^b} &= m^{ab}.
\end{align}
To consider gauging by the Courant algebroid $E$,
a connection $\nabla$ on $E$ is introduced. 
Throughout this paper, assume the connection $\nabla$ is 
a metric connection, i.e., $\nabla m = 0$.
Let $\omega = \omega^b_{ai} \rd x^i \otimes  e^a \otimes e_b$ 
be its connection $1$-form
satisfying $\nabla e_a = - \omega_{ai}^b \rd x^i \otimes e_b$.

Similar to the case of Lie algebroid gauging in Section \ref{sec:SCSMLA},
$z_i$ is not transformed covariantly.
From coordinate transformation formulas of graded manifolds 
in Section \ref{manicovariant}, the covariantized coordinate is 
$z_i^{\nabla} := z_i + \frac{1}{2} \omega_{bi}^e m_{ec} a^b a^c
= z + \frac{1}{2} \omega_{bci} a^b a^c$.

The Liouville $1$-form such that $\gomega = - \delta \vartheta$ 
is covariantized as
\begin{align}
\vartheta &= 
z_i \delta x^i 
- p_i \delta q^i 
- \frac{1}{2} m_{ab} a^a \delta a^b
\nonumber \\
& = z_i^{\nabla} \delta x^i 
- p_i \delta q^i 
- \frac{1}{2} m_{ab} a^a D a^b,
\label{Liouville4}
\end{align}
which decides the kinetic term of the AKSZ action,
where $D a^a = \delta a^a - \omega_{bi}^a a^b \delta x^i$.
Note that $\vartheta^{\nabla} = \vartheta$ since 
the Liouville $1$-form is invariant under local (graded) diffeomorphisms
of the graded manifold.
Nontrivial Poisson brackets including $z_i^{\nabla}$ are
\begin{align}
\sbv{z_i^{\nabla}}{a^a} &= \omega_{ai}^b a^b,
\\
\sbv{z_i^{\nabla}}{z_j^{\nabla}} 
&= \frac{1}{2} R_{ijb}^a m_{ac} a^c a^b 
= \frac{1}{2} R_{ijab} a^a a^b.
\end{align}
Note that $\sbv{x^i}{z_j^{\nabla}} = \delta^i_j$.
where $R$ is a curvature in Eq.~\eqref{curv2} or Eq.~\eqref{curvlocal2}.

For an AKSZ sigma model, we start at homological functions 
$\Thetas$ for the standard Courant algebroid
and $\Thetae$ for a general Courant algebroid,
\begin{align}
\Thetas &= z_i q^i + \frac{1}{3!} H_{ijk}(x) q^i q^j q^k,
\\
\Thetae &= \rho^i_a(x) z_i a^a + 
\frac{1}{3!} C_{abc}(x) a^a a^b a^c,
\end{align}
$\Thetas$ is not invariant under local coordinate transformations 
on $T^*[2]E[1]$ since $z_i$ is not a covariant coordinate.
'Gauging' requires invariance of homological functions under
diffeomorphism on $T^*[2]E[1]$.
Invariant homological functions are obtained by 
replacing $z_i$ to $z_i^{\nabla}$ as
\begin{align}
\Thetas^{\nabla} &= \Thetas
+ \frac{1}{2} \omega_{ai}^b a^a m_{bc} a^c q^i
= z_i^{\nabla} q^i 
+ \frac{1}{3!} H_{ijk}(x) q^i q^j q^k.
\label{SCSMCAhomological01}
\end{align}
On the other hand, $\Thetae$ is already invariant since it is rewritten 
to the manifest covariant expression as
\begin{align}
\Thetae^{\nabla} &= \Thetae = \rho^i_a(x) z_i^{\nabla} a^a + 
\frac{1}{3!} T_{abc}(x) a^a a^b a^c,
\label{SCSMCAhomological02}
\end{align}
where $T_{abc}$ is the $E$-torsion, or the Gualtieri torsion on $E$, 
Eq.\eqref{Etorsion2}.
Poisson brackets of $\Thetas^{\nabla}$ and $\Thetae^{\nabla} = \Thetae$
are 
\begin{align}
\sbv{\Thetas^{\nabla}}{\Thetas^{\nabla}} 
& = \frac{1}{2} R_{ijd}^c m_{ce} q^i q^j a^e a^d,
\label{TT01}
\\
\sbv{\Thetae}{\Thetae} &= 0,
\label{TT02}
\\
\sbv{\Thetas^{\nabla}}{\Thetae} 
&= - \nabla_i \rho^j_a z_j^{\nabla} q^i a^a 
- \frac{1}{3!} S_{iabc} q^i a^a a^b a^c
+ \frac{1}{3!} \rho^l_{a} \partial_{l} H_{ijk} q^i q^j q^k a^a,
\label{TT03}
\end{align}
where $R_{ijd}^c$ is the curvature of an ordinary connection, 
Eq.~\eqref{curvlocal2}, 
$S_{iabc}$ is the basic curvature of the Courant algebroid, Eq.~\eqref{bcurvlocal2}.

From Eqs.~\eqref{TT01}--\eqref{TT03}, we obtain the homological condition 
satisfying $\sbv{\Theta^{\nabla}}{\Theta^{\nabla}}=0$.
The total function $\Theta^{\nabla} = \Thetas^{\nabla} + \Thetae$ is 
homological if and only if
\begin{align}
& R_{ija}^b =0,
\\
& S_{iabc} = 0,
\\
& \nabla_i \rho^j_a = 0,
\label{homological13}
\\
& \rho^l_a(x) \partial_{l} H_{ijk} = 0.
\end{align}
Similar to Lie algebroid gauging case in Section \ref{sec:SCSMLA},
these identities give geometric conditions to make $\Theta^{\nabla}$
homological.
\begin{theorem}
$\Theta^{\nabla}$ is homological if and only if
$R = {}^E S =0$,
$\nabla \rho = 0$,
and ${}^E \nabla H =0$.
\end{theorem}
$Q^{\nabla} = \sbv{\Theta^{\nabla}}{-}$ can be regarded as 
an equivariant differential if we do not impose the above 
'flatness' conditions as in Lie algebroid gauging cases.

The AKSZ sigma model is constructed from the above homological function 
$\Theta^{\nabla}$. The AKSZ-BV action functional is
\begin{align}
S &= S_S^{\nabla} + S_E
\nonumber \\ 
&=\int_{T[1]\Sigma} \rd^3 \sigma \rd^3 \theta
\left[- \bz_i ^{\nabla} \rd \bX^i 
+ \bp_i \rd \bq^i 
+ \frac{1}{2} m_{ab} \ba^a D \ba^b
+ \bz_i ^{\nabla} \bq^i
+ \rho_a^i \bz_i^{\nabla} \ba^a
\right.
\nonumber \\
& 
\left.
+ \frac{1}{3!} H_{ijk}(\bX) \bq^i \bq^j \bq^k 
+ \frac{1}{3!} T_{abc}(\bX) \ba^a \ba^b \ba^c \right],
\label{SCSMCAaction}
\end{align}
As in Section \ref{sec:SCSMLAflux}, we can introduce flux terms to consider 
more general gauged actions. Such a generalization is discussed 
in Section \ref{sec:SCSMCAflux}.

\subsection{General Courant sigma models with Lie algebroid gauging}\label{sec:GCSMLA}
As a total structure, we take a Courant sigma model with a general 
Courant algebroid $E$ instead of $TM \oplus T^*M$.
In this section, we consider gauging by a Lie algebroid $A$.
If a Lie algebroid is an action Lie algebroid $A = M \times \mathfrak{g}$,
it is ordinary gauging by a Lie group $G$ with taking $\mathfrak{g}$ as 
a Lie algebra of $G$.

A typical and interesting example of this setting is 
that $A$ is a Dirac subbundle of $E$
since a Dirac structures is a Lie algebroid itself.
However we do not assume that $A$ is a subbundle of $E$,
nor any relation between $E$ and $A$ except that
both bundles have the same base manifold $M$.
Note that for a Courant algebroid $E$ and a Lie algebroid $A$, 
$E \oplus A$ is not necessarily a Courant algebroid.

We consider the graded manifold $T^*[2](E[1] \oplus A[1])$ for construction.
Take Local coordinates $(x^i, \eta^A, a^a)$ on $E[1] \oplus A[1]$
of degree $(0, 1, 1)$, where $x^i$ is a coordinate on $M$, $\eta^A$ is 
a fiber coordinate on 
$E[1]$ and  $a^a$ is a fiber coordinate on $A[1]$,
Here $i = 1, \ldots, \dim(M) $, $A = 1, \ldots, \mathrm{rank}(E) $ and 
$a = 1, \ldots, \mathrm{rank}(A) $.
Canonical conjugate coordinates are denoted by $(z_i, k_{AB} \eta^B, w_a)$ of degree 
$(2, 1, 1)$, where $k_{AB}$ is a fiber metric on $E$, which gives
the inner product $\bracket{-}{-}$ of the Courant algebroid $E$.
The graded symplectic form is
\begin{align}
\gomega &= 
\delta x^i \wedge \delta z_i 
+ \frac{1}{2} \delta \eta^A \wedge \delta (k_{AB} \eta^B)
+ \delta a^a \wedge \delta w_a
\label{gsymplectic5}
\end{align}
Canonical Poisson brackets induced from $\gomega$ are
\begin{align}
\sbv{x^i}{z_j} &= \delta^i_j,
\\
\sbv{\eta^A}{\eta^B} &= k^{AB},
\\
\sbv{a^a}{w_b} &= \delta^a_b,
\end{align}
We rewrite local coordinates to covariant forms with respect to 
coordinate transformations of $T^*[2]A[1]$
by introducing an ordinary vector bundle connection $\nabla$ on $A$. 
Let $\omega = \omega_{ai}^b \rd x^i \otimes e_b \otimes e^a$
be its connection $1$-form.
Assume that $\nabla$ is a metric connection, satisfying $\nabla k = 0$.
The covariant coordinate is 
$z_i^{\nabla} := z_i + \omega_{bi}^c a^b w_c$
introduced in Section \ref{manicovariant}.
Note that we can also introduce an ordinary connection $\nabla^{\prime}$ on $E$
and the connection $1$-form 
$\Omega = \Omega_{Ai}^B \rd x^i \otimes e_B \otimes e^A$.
Then the totally covariant coordinate on $T^*[2](E[1] \oplus A[1])$ becomes 
$z_i^{\nabla} := z_i - \frac{1}{2} \Omega_{Bi}^D k_{DC} \eta^B \eta^C 
+ \omega_{bi}^c a^b w_c$.
The total action of the gauged CSM $S$ depends on $\omega_{ai}^b$ 
but does not depend on $\Omega_{Ai}^B$.
Thus we do not introduce a connection on $E$ to simplify discussions and 
concentrate on analysis of $\omega_{ai}^b$ dependence, i.e., a connection on 
$A$, in the action functional $S$. 

The Liouville form such that $\gomega = - \delta \vartheta$ is
\begin{align}
\vartheta &= 
z_i \delta x^i 
- \frac{1}{2} k_{AB} \eta^A \delta \eta^B 
- w_a \delta a^a 
\nonumber \\
& = z_i^{\nabla} \delta x^i 
- \frac{1}{2} k_{AB} \eta^A \delta \eta^B 
- w_a D a^a,
\label{Liouville5}
\end{align}
which in fact invariant under the changing of local coordinates
\eqref{coordinatetransf1}--\eqref{coordinatetransf3},
where $D a^a = \delta a^a - \omega^a_{bi} a^b \delta x^i$,

Poisson brackets including $z_i^{\nabla}$ are 
\begin{align}
\sbv{x^i}{z_j^{\nabla}} &= \delta^i_j,
\\
\sbv{z_i^{\nabla}}{a^a} &= \omega_{bi}^a a^b,
\\
\sbv{z_i^{\nabla}}{w_a} &= - \omega_{ai}^b w_b,
\\
\sbv{z_i^{\nabla}}{z_j^{\nabla}} &= R_{ijb}^a w_a a^b,
\end{align}

In order to construct the AKSZ action, we start at homological functions 
of a Courant algebroid $E$ and a Lie algebroid $A$,
\begin{align}
\Thetac &= \rhoe^i_A(x) z_i \eta^A + \frac{1}{3!} H_{ABC}(x) \eta^A \eta^B \eta^C,
\\
\Thetaa &= \rhoa^i_a(x) z_i a^a + 
\frac{1}{2} C_{ab}^c(x) a^a a^b w_c,
\end{align}
where $\rhoe:E \rightarrow TM$ is the anchor map of the Courant algebroid 
$E$ and 
$\rhoa: A \rightarrow TM$ is the anchor map of the Lie algebroid $A$.
$\Thetac$ is not invariant under local coordinate transformations on $T^*[2]A[1]$.
On the other hand, $\Thetaa$ is invariant.
In order to make $\Thetac$ covariant, $z_i$ is replaced to $z^{\nabla}_i$.
Covariantized homological functions are
\begin{align}
\Thetac^{\nabla} &= \Thetac
+ \rhoe^i_A \omega_{bi}^c a^b w_c \eta^A
= \rhoe^i_A(x) z^{\nabla}_i \eta^A + \frac{1}{3!} H_{ABC}(x) \eta^A \eta^B \eta^C,
\label{GCSMLAhomological01}
\\
\Thetaa^{\nabla} &= \Thetaa = \rho^i_a(x) z_i^{\nabla} a^a 
- \frac{1}{2} T_{ab}^c(x) a^a a^b w_c,
\label{GCSMLAhomological02}
\end{align}
where $T_{ab}^c$ is the $A$-torsion in the Lie algebroid $A$.
Since $\Thetac$ is not equal to $\Thetac^{\nabla}$, 
$\Thetac^{\nabla}$ does not satisfies the homological condition 
$\sbv{\Theta}{\Theta}=0$ any more without extra conditions.
On the other hand, similar to Section \eqref{sec:SCSMLA},
$\Thetaa^{\nabla} = \Thetaa$ is not deformed.
Poisson brackets of  $\Thetac^{\nabla}$ and $\Thetaa$ are
\begin{align}
\sbv{\Thetac^{\nabla}}{\Thetac^{\nabla}} 
& = - \rhoe^i_A \rhoe^j_B R_{ijc}^d a^c w_d \eta^A \eta^B,
\label{TTT01}
\\
\sbv{\Thetaa}{\Thetaa} &= 0,
\label{TTT02}
\\
\sbv{\Thetac^{\nabla}}{\Thetaa} 
&= (\rhoe^j_A \nabla_j \rhoa^i_a - \rhoa^j_a \partial_j \rhoe^i_A)
z_i^{\nabla} a^a \eta^A 
\nonumber \\ & \qquad
- \frac{1}{2} \rhoe^i_A S_{iab}^c a^a a^b w_c \eta^A
- \frac{1}{3!} \rhoa^l_{a} \partial_{l} H_{ABC} a^a \eta^A \eta^B \eta^C.
\label{TTT03}
\end{align}
Thus, for $\Theta^{\nabla} = \Thetac^{\nabla} + \Thetaa$,
$\Theta^{\nabla}$ is homological, i.e.,
$\sbv{\Theta^{\nabla}}{\Theta^{\nabla}}=0$ if and only if
\begin{align}
& R_{ija}^b =0,
\label{GCSMLAcondition01}
\\
& \rhoe^j_A \nabla_j \rhoa^i_a - \rhoa^j_a \partial_j \rhoe^i_A =0,
\\
& \rhoe^i_A S_{iab}^c =0,
\\
& \rhoa^l_{a} \partial_{l} H_{ABC} = 0.
\label{GCSMLAcondition04}
\end{align}
Similar to previous sections
these conditions give geometric conditions 
to the target space.
\begin{theorem}\label{CAwithLA}
$\Theta^{\nabla}$ is homological if and only if
$R=0$,  $\iota_{\rhoe} {}^A S = 0$,
${}^A \nabla \rhoe = 0$ and ${}^A \rd H = 0$.
Then, for the Courant algebroid $E$ and the Lie algebroid $A$,
$T^*[2](E[1] \oplus A[1])$ becomes a QP-manifold.
\end{theorem}

An AKSZ sigma model is constructed from above data.
The AKSZ action functional is denoted by $S = S_C + S_A$.
$S_C$ is constructed from the covariantized deformation, 
Eq.~\eqref{CSMAKSZaciton} and the kinetic term induced from 
Eq.~\eqref{Liouville5}.
The covariantized action is
\begin{align}
S_C^{\nabla} &= \int_{T[1]\Sigma} \rd^3 \sigma \rd^3 \theta
\left[- \bz_i^{\nabla} \rd \bX^i 
+ \frac{1}{2} k_{AB} \by^A \rd \by^B
+ \rhoe^i_A \bz_i^{\nabla} \by^A
\right.
\nonumber \\ &
\left.
\qquad + \frac{1}{3!} H_{ABC}(\bX) \by^A \by^B \by^C \right],
\end{align}
where $\bz_i^{\nabla} = \bz_i + \omega_{bi}^c \ba^b \bw_c$.
$S_A$ is induced from $\Thetaa$ with a kinetic term,
\begin{align}
S_A &= \int_{T[1]\Sigma} \rd^3 \sigma \rd^3 \theta
\left[\bw_a D \ba^a
+ \rhoa^i_a(\bX) \bz_i \ba^a 
+ \frac{1}{2} C_{ab}^c \ba^a \ba^b \bw_c \right]
\nonumber \\
&= \int_{T[1]\Sigma} \rd^3 \sigma \rd^3 \theta
\left[\bw_a D \ba^a
+ \rho^i_a(\bX) \bz_i^{\nabla} \ba^a 
- \frac{1}{2} T_{ab}^c \ba^a \ba^b \bw_c \right],
\end{align}
where $D \ba^a = \rd \ba^a - \omega^a_{bi} \ba^b \rd \bX^i$.
The total action functional is
\begin{align}
S &= S_C^{\nabla} + S_A
\nonumber \\ 
&=\int_{T[1]\Sigma} \rd^3 \sigma \rd^3 \theta
\left[- \bz_i^{\nabla} \rd \bX^i 
+ \frac{1}{2} k_{AB} \by^A \rd \by^B
+ \bw_a D \ba^a
+ \rhoe^i_A \bz_i^{\nabla} \by^A
\right.
\nonumber \\
& 
\left.
\quad 
+ \frac{1}{3!} H_{ABC}(\bX) \by^A \by^B \by^C 
+ \rhoa^i_a(\bX) \bz_i^{\nabla} \ba^a 
- \frac{1}{2} T_{ab}^c(\bX) \ba^a \ba^b \bw_c \right].
\label{GCSMLAaction}
\end{align}

\subsection{General Courant sigma models with Courant algebroid gauging}\label{sec:GCSMCA}
In this section, we consider a Courant sigma model 
such that the target space is a general Courant algebroid $E$, 
and gauging by another general Courant algebroid $A$.
In a typical case $A$ is a Courant subalgebroid of $E$, 
however we do not assume any relation between $E$ and $A$ for construction.

First we construct a QP-manifold as a target space structure. 
Take the graded manifold $T^*[2](E[1] \oplus A[1])$,
and take Local coordinates $(x^i, \eta^A, a^a)$ of degree 
$(0, 1, 1)$, where $x^i$ is a coordinate on $M$, $\eta^A$ is a fiber coordinate on $E[1]$ and $a^a$ is a fiber coordinate on $A[1]$. Here $i = 1, \ldots, \dim(M) $, $A = 1, \ldots, \mathrm{rank}(E) $ and $a = 1, \ldots, \mathrm{rank}(A)$.

Canonical conjugate coordinates are denoted by 
$(z_i, k_{AB} \eta^B, m_{ab} a^b)$ 
of degree $(2, 1, 1)$, where $k_{AB}$ is a fiber metric on $E$ and
$m_{ab}$ is a fiber metric on $A$, 
which are the inner products of each Courant algebroid.

Introduce a connection $\nabla$ on $A$. 
Throughout this paper, assume the connection $\nabla$ is a metric connection,
i.e., $\nabla k = \nabla m = 0$.

We take the 'canonical' graded symplectic form,
\begin{align}
\gomega &= 
\delta x^i \wedge \delta z_i 
+ \frac{1}{2} \delta \eta^A \wedge \delta (k_{AB} \eta^B)
+ \frac{1}{2} \delta a^a \wedge \delta (m_{ab} a^b)
\label{gsymplectic6}
\end{align}
Then, nontrivial Poisson brackets are given by
\begin{align}
\sbv{x^i}{z_j} &= \delta^i_j,
\\
\sbv{\eta^A}{\eta^B} &= k^{AB},
\\
\sbv{a^a}{a^b} &= m^{ab},
\end{align}

$z_i$ is covariantized to the covariant local coordinate
$z_i^{\nabla} := z_i + \frac{1}{2} \omega_{ai}^c m_{cb} a^a a^b$.
The Liouville $1$-form such that $\gomega = - \delta \vartheta$ is
\begin{align}
\vartheta &= 
z_i \delta x^i 
- \frac{1}{2} k_{AB} \eta^A \delta \eta^B 
- \frac{1}{2} m_{ab} a^a \delta a^b 
\nonumber \\
& = z_i^{\nabla} \delta x^i 
- \frac{1}{2} k_{AB} \eta^A \delta \eta^B 
- \frac{1}{2} m_{ab} a^a D a^b,
\label{Liouville6}
\end{align}
where $D a^a = \delta a^a - \omega^a_{bi} a^b \delta x^i$.
$\vartheta$ is invariant under changing of local coordinates 
on the graded manifold $T^*[2](E[1] \oplus A[1])$.
Here transformations of local coordinates on a graded manifold 
$T^*[2]A[1]$ is given 
by Eqs.~\eqref{coordinatetransf1}--\eqref{coordinatetransf3}
for corresponding coordinates.
Note that we do not consider covariant coordinate with respect to 
$T^*[2]E[1]$ by introducing a connection on $E$, 
since geometric structures of the sigma model, especially homological functions
do not depend on the connection on $E$.
One can refer to Appendix \eqref{connectionCA} for curvatures, $E$-connections, $E$-torsions, basic curvatures, etc. in Courant algebroids.

Nonzero Poisson brackets including $z_i^{\nabla}$ are
\begin{align}
\sbv{x^i}{z_j^{\nabla}} &= \delta^i_j,
\\
\sbv{z_i^{\nabla}}{a^a} &= \omega_{bi}^a a^b,
\\
\sbv{z_i^{\nabla}}{z_j^{\nabla}} &= 
\frac{1}{2} R_{ijb}^d m_{da} a^a a^b.
\end{align}
Homological functions of Courant algebroids $E$ and $A$ are
\begin{align}
\Thetac &= \rhoe^i_A(x) z_i \eta^A + \frac{1}{3!} H_{ABC}(x) \eta^A \eta^B \eta^C,
\\
\Thetaa &= \rhoa^i_a(x) z_i a^a + \frac{1}{3!} C_{abc}(x) a^a a^b a^c.
\end{align}
$\Thetaa$ is an invariant function since 
the expression is rearranged to manifestly covariant form,
$\Thetaa = \rho^i_a(x) z_i^{\nabla} a^a 
+ \frac{1}{3!} T_{abc}(x) a^a a^b a^c$,
where $T_{abc}$ is the $A$-torsion for the Courant algebroid $A$.
However $\Thetac$ is not invariant.
We deform $\Thetac$ by changing $z_i$ to $z^{\nabla}_i$,
and make $\Thetac$ invariant under local coordinate transformations 
on $T^*[2]A[1]$.
Then, covariantized homological functions are
\begin{align}
\Thetac^{\nabla} &= \Thetac
+ \frac{1}{2} \rhoe^i_A \omega_{bi}^d m_{dc} a^b a^c \eta^A
= \rhoe^i_A(x) z^{\nabla}_i \eta^A + \frac{1}{3!} H_{ABC}(x) \eta^A \eta^B \eta^C,
\label{homologicalCA1}
\\
\Thetaa^{\nabla} &= \Thetaa = \rhoa^i_a(x) z_i^{\nabla} a^a 
+ \frac{1}{3!} T_{abc}(x) a^a a^b a^c,
\label{homologicalCA2}
\end{align}
Their Poisson brackets are as follows.
\begin{align}
\sbv{\Thetac^{\nabla}}{\Thetac^{\nabla}} 
& = \frac{1}{2} \rhoe^i_A \rhoe^j_B R_{ijc}^e m_{ed} a^c a^d \eta^A \eta^B
+ \rhoe^i_A \rhoe^j_B \omega_{aj}^e \nabla_i m_{ed} a^c a^d \eta^A \eta^B,
\\
\sbv{\Thetaa}{\Thetaa} &= 0,
\\
\sbv{\Thetac^{\nabla}}{\Thetaa} 
&= (\rhoe^j_A \nabla_j \rhoa^i_a - \rhoa^j_a \partial_j \rhoe^i_A)
z_i^{\nabla} a^a \eta^A 
+ \frac{1}{3!} \rhoe^i_A S_{iabc} a^a a^b a^c \eta^A
\nonumber \\ & \qquad 
+ \frac{1}{3!} \rhoa^l_{a} \partial_{l} H_{ABC} a^a \eta^A \eta^B \eta^C.
\end{align}
Therefore, the total function $\Theta^{\nabla} = \Thetac^{\nabla} + \Thetaa$
satisfies the homological condition $\sbv{\Theta^{\nabla}}{\Theta^{\nabla}}=0$
if and only if the following identities satisfied,
\begin{align}
& R_{ija}^b =0,
\\
& \rhoe^i_A S_{iabc} =0,
\\
& \rhoe^j_A \nabla_j \rhoa^i_a - \rhoa^j_a \partial_j \rhoe^i_A =0,
\\
& \rhoa^l_{a} \partial_{l} H_{ABC} = 0.
\end{align}
Similar to Lie algebroid gauging case in Section \ref{sec:GCSMLA},
these conditions give geometric conditions to make $\Theta^{\nabla}$
homological.
\begin{theorem}\label{CAwithCA}
$\Theta^{\nabla}$ is homological if and only if
$R=0$,  $\iota_{\rhoe} {}^A S = 0$,
${}^A \nabla \rhoe = 0$ 
and ${}^A \rd H = 0$.
Then, for the Courant algebroid $E$ and another Courant algebroid $A$,
$T^*[2](E[1] \oplus A[1])$ is a QP-manifold.
\end{theorem}
$Q^{\nabla} = \sbv{\Theta^{\nabla}}{-}$ can be regarded as 
an equivariant differential if we do not impose 'flatness' conditions.

Settings and results in Theorem \ref{CAwithLA} and 
\ref{CAwithCA} are geometric results connected to analysis of 
symmetries and reductions of Courant algebroids
in \cite{BursztynCavalcantiGualtieri, BursztynCavalcantiGualtieri2,
BursztynIglesiasPonteSevera, Bursztyn:2023onv}. 
Especially, \cite{Bursztyn:2023onv} uses graded geometry.
We obtain QP-manifold and AKSZ sigma model realizations.

The AKSZ sigma model is constructed from the homological function 
$\Theta^{\nabla}$. 
Note that from the Liouville $1$-form \eqref{Liouville6},
the kinetic term is already invariant,
\begin{align}
& \int_{T[1]\Sigma} \rd^3 \sigma \rd^3 \theta
\left[- \bz_i \rd \bX^i 
+ \frac{1}{2} k_{AB} \by^A \rd \by^B
+ \frac{1}{2} m_{ab} \ba^a \rd \ba^b
\right]
\nonumber \\ &
= \int_{T[1]\Sigma} \rd^3 \sigma \rd^3 \theta
\left[- \bz_i^{\nabla} \rd \bX^i 
+ \frac{1}{2} k_{AB} \by^A \rd \by^B
+ \frac{1}{2} m_{ab} \ba^a D \ba^b
\right]
\end{align}
By deforming the action functional of the original CSM 
Eq.~\eqref{CSMAKSZaciton},
the covariantized action is constructed from 
Eq.~\eqref{homologicalCA1},
\begin{align}
S_C^{\nabla} &= \int_{T[1]\Sigma} \rd^3 \sigma \rd^3 \theta
\left[- \bz_i^{\nabla} \rd \bX^i 
+ \frac{1}{2} k_{AB} \by^A \rd \by^B
+ \rhoe^i_A \bz_i^{\nabla} \by^A
\right.
\nonumber \\ &
\left.
\qquad + \frac{1}{3!} H_{ABC}(\bX) \by^A \by^B \by^C \right],
\end{align}
where $\bz_i^{\nabla} = \bz_i - \frac{1}{2} \omega_{bi}^d m_{dc} \ba^b \ba^c$.
Gauging terms including $\ba^a$ are induced from 
Eq.~\eqref{homologicalCA2} as
\begin{align}
S_A &= \int_{T[1]\Sigma} \rd^3 \sigma \rd^3 \theta
\left[\frac{1}{2} m_{ab}(\bX) \ba^a D \ba^b
+ \rhoa^i_a(\bX) \bz_i \ba^a 
+ \frac{1}{3!} C_{abc}(\bX) \ba^a \ba^b \ba^c \right]
\nonumber \\
&= \int_{T[1]\Sigma} \rd^3 \sigma \rd^3 \theta
\left[\frac{1}{2} m_{ab}(\bX) \ba^a D \ba^b
+ \rhoa^i_a(\bX) \bz_i^{\nabla} \ba^a 
+ \frac{1}{3!} T_{abc}(\bX) \ba^a \ba^b \ba^c \right]
\end{align}
where $D \ba^a = \rd \ba^a - \omega^a_{bi} \ba^b \rd \bX^i$.
Therefore, the total action functional of the GCSM is
\begin{align}
S &= S_C^{\nabla} + S_A
\nonumber \\ 
&=\int_{T[1]\Sigma} \rd^3 \sigma \rd^3 \theta
\left[- \bz_i^{\nabla} \rd \bX^i 
+ \frac{1}{2} k_{AB}(\bX) \by^A \rd \by^B
+ \frac{1}{2} m_{ab}(\bX) \ba^a D \ba^b
\right.
\nonumber \\
& 
\left.
+ \rhoe^i_A \bz_i^{\nabla} \by^A
+ \frac{1}{3!} H_{ABC}(\bX) \by^A \by^B \by^C
+ \rhoa^i_a(\bX) \bz_i^{\nabla} \ba^a 
+ \frac{1}{3!} T_{abc}(\bX) \ba^a \ba^b \ba^c \right].
\label{GCSMCAaction}
\end{align}

\section{SCSMs with fluxes and boundaries}\label{sec:CSMflux}
\noindent
In this section, we consider extensions of the gauged Courant sigma model 
by introducing flux terms and boundary terms.
In this section, we concentrate on the SCSM with Lie algebroid gauging
in Section \ref{sec:SCSMCA}.
Other cases, the SCSM with Courant algebroid gauging, 
and the general CSM with Lie algebroid and Courant algebroid gauging
are analyzed in Appendix \ref{sec:SCSMCAflux}, \ref{sec:GCSMLAflux}
and \ref{sec:GCSMCAflux}, which 
are generalizations of the model in this section. 

\subsection{Introduction of fluxes}\label{sec:SCSMLAflux}
The homological function
$\Theta^{\nabla} := \Thetas^{\nabla} + \Thetaa^{\nabla}$, 
with Eq.~\eqref{thetas01} and \eqref{thetaa01} is not most general.
More terms of degree three can be added to $\Theta^{\nabla}$.
Here we add the following terms,
\begin{align}
\Theta_F &=
\frac{1}{2} H^{(2)}_{ija}(x) q^i q^j a^a 
+ \frac{1}{2} H^{(1)}_{iab}(x) q^i a^a a^b
+ \frac{1}{3!} H^{(0)}_{abc}(x) a^a a^b a^c.
\label{thetaf}
\end{align}
Here $H^{(2)} = \frac{1}{2} H^{(2)}_{ija}(x) \rd x^i \wedge \rd x^j \otimes e^a
\in \Omega^2(M, A^*)$ 
is a two-form taking a value in $A^*$,
$H^{(1)} = \frac{1}{2} H^{(1)}_{iab}(x) \rd x^i \otimes e^a \wedge e^b.
\in \Omega^1(M, \wedge^2 A^*)$
is a one-form taking a value in $\wedge^2 A^*$,
and $H^{(0)}= \frac{1}{3!} H^{(0)}_{abc}(x) e^a \wedge e^b \wedge e^c.
\in \Omega^0(M, \wedge^3 A^*)$
is a zero-form taking a value in $\wedge^3 A^*$.

More generally, degree three terms in a function $H(x, q, p, a, w)$ 
can be added in the homological function.
Since $q, p, a, w$ are degree coordinates, they are third order functions of 
$q, p, a, w$.
They are called fluxes in physical contexts to deform original physical theories.
Then, the total homological function is 
$\Theta^{\nabla} := \Thetas^{\nabla} + \Thetaa^{\nabla} + \Theta_F$.
By deforming the homological function, consistency conditions are deformed.

New Poisson brackets including $\Thetaf$ are as follows,
\footnote{Notation: $\partial_{[i} \alpha_{j]}
= \partial_{i} \alpha_{j}- \partial_{j} \alpha_{i}$,
and 
$\partial_{[i} T_{jk]}
= \partial_{i} T_{jk} + \partial_{j} T_{ki} + \partial_{k} T_{ij}$
for an antisymmetric tensor $T_{ij}$, etc. }
\begin{align}
\sbv{\Thetas^{\nabla}}{\Thetaf} 
& = - \frac{1}{3!} \nabla_{i} H^{(2)}_{jka}(x) q^i q^j q^k a^a 
+ \frac{1}{4} \nabla_{i}H^{(1)}_{jab}(x) q^i q^j a^a a^b
- \frac{1}{3!} \partial_{i}H^{(0)}_{abc}(x) q^i a^a a^b a^c.
\label{cme04}
\\
\sbv{\Thetaa^{\nabla}}{\Thetaf} 
& = 
- \frac{1}{4} (\rho^i_{a} \partial_i H^{(2)}_{ijb}
- C_{ab}^c H^{(2)}_{ijc} ) q^i q^j a^a a^b
+ \frac{1}{3!} (\rho^i_{a} \partial_i H^{(1)}_{ibc} 
- C_{ab}^d H^{(1)}_{idc} )q^i a^a a^b a^c
\nonumber \\ & \quad
- \frac{1}{4!} (\rho^i_{a} \partial_i H^{(0)}_{bcd} 
- C_{ab}^e H^{(0)}_{ecd}) a^a a^b a^c a^d
\nonumber \\
& = 
- \frac{1}{4} (\rho^k_a \partial_k H^{(2)}_{ijb}
- C_{ab}^c H^{(2)}_{ijc} ) q^i q^j a^a a^b
+ \frac{1}{3!} (\rho^i_{a} \partial_k H^{(1)}_{ibc} 
- C_{ab}^d H^{(1)}_{idc} )q^i a^a a^b a^c
\nonumber \\ & \quad
- \frac{1}{4!} (\rho^i_{a} \partial_k H^{(0)}_{bcd} 
- C_{ab}^e H^{(0)}_{ecd}) a^a a^b a^c a^d.
\label{cme05} \\
\sbv{\Thetaf}{\Thetaf}  
& = 0.
\label{cme06}
\end{align}
$\sbv{\Theta^{\nabla}}{\Theta^{\nabla}} =0$ is imposed for 
the deformed function,
$\Theta^{\nabla} = \Thetas^{\nabla} + \Thetaa^{\nabla} + \Theta_F$.
Fluxes must satisfy some geometric conditions 
obtained from Eqs~.\eqref{cme04}-\eqref{cme06}.
Combining them
with \eqref{cme01}-\eqref{cme03},
we obtain conditions.
They are as follows,
\begin{align}
& R_{ija}^b =0,
\\
& S_{iab}^c =0,
\\
& \nabla_i \rho^j_a =0,
\\
& - \nabla_{[i} H^{(2)}_{jk]a}(x)
+ {}^A \rd_{a} H_{ijk}(x)  = 0,
\\
& - \nabla_{[i}H^{(1)}_{j]ab}(x) 
+ {}^A \rd_{[a} H^{(2)}_{ij|b]}(x) = 0,
\\
& - \partial_{i}H^{(0)}_{abc}(x)
+ {}^A \rd_{[a|}H^{(1)}_{i|bc]}(x) = 0,
\\
& {}^A \rd_{[a}H^{(0)}_{bcd]}(x) = 0.
\end{align}
Here the Lie algebroid differential is ${}^A \rd_{[a} H^{(2)}_{ij|b]}
= \rho^i_{[a} \partial_i H^{(2)}_{ij|b]}
- C_{ab}^c H^{(2)}_{ijc}$, etc.
Note that ${}^A \rd = {}^A \rd^{\nabla}$ under $\nabla \rho = 0$
for the basic $A$-connection.
Therefore, we obtain the following result.
\begin{theorem}\label{SCSMLAflux}
$\Theta^{\nabla} = \Thetas + \Thetaa + \Thetaf$.
is homological if and only if the following conditions are satisfied,
\begin{align}
& R = {}^A S =0,  \quad \nabla \rho = 0,
\label{fluxeq01}
\\
& {}^A \rd^{\nabla} H - \nabla H^{(2)} = 0,
\label{fluxeq02}
\\
& {}^A \rd^{\nabla} H^{(2)} - \nabla H^{(1)} = 0,
\label{fluxeq03}
\\
& {}^A \rd^{\nabla} H^{(1)} - \rd H^{(0)} = 0,
\label{fluxeq04}
\\
& {}^A \rd^{\nabla} H^{(0)} =0.
\label{fluxeq05}
\end{align}

\end{theorem}
One of nontrivial solutions of Eqs.~\eqref{fluxeq01}--\eqref{fluxeq05}
is as follows.
For a given closed $3$-form $H \in \Omega^3(M)$,
$H^{(2)} = \iota_{\rho} H$, 
$H^{(1)} = \iota_{\rho} \iota_{\rho} H$, 
and $H^{(0)} = \iota_{\rho} \iota_{\rho} \iota_{\rho} H$,
where $\iota_{\rho}(e_3) \iota_{\rho}(e_2) \iota_{\rho}(e_1) H
:= H(\rho(e_1), \rho(e_2), \rho(e_3))$ for $e_1, e_2, e_3 \in \Gamma(A)$, 
etc.

In the AKSZ action \eqref{actionSCSMLA},
the function $\Thetaf$, Eq.~\eqref{thetaf} corresponds to the following term,
\begin{align}
S_F
&= \int_{T[1]\Sigma} \rd^3 \sigma \rd^3 \theta
\left[\frac{1}{2} H^{(2)}_{ija}(\bX) \bq^i \bq^j \ba^a 
+ \frac{1}{2} H^{(1)}_{iab}(\bX) \bq^i \ba^a \ba^b
+ \frac{1}{3!} H^{(0)}_{abc}(\bX) \ba^a \ba^b \ba^c
\right].
\label{SCSMLAfluxaction}
\end{align}
Therefore the total action is 
\begin{align}
S &= S_S^{\nabla} + S_A + S_F
\nonumber \\ 
&=\int_{T[1]\Sigma} \rd^3 \sigma \rd^3 \theta
\left[- \bz_i ^{\nabla} \rd \bX^i 
+ \bp_i \rd \bq^i 
+ \bw_a D \ba^a
+ \bz_i^{\nabla} \bq^i
+ \rho^i_a(\bX) \bz_i^{\nabla} \ba^a 
\right.
\nonumber \\
& 
\left.
+ \frac{1}{3!} H^{(3)}_{ijk}(\bX) \bq^i \bq^j \bq^k 
- \frac{1}{2} T_{ab}^c(\bX) \ba^a \ba^b \bw_c 
\right.
\nonumber \\
& 
\left.
+ \frac{1}{2} H^{(2)}_{ija}(\bX) \bq^i \bq^j \ba^a 
+ \frac{1}{2} H^{(1)}_{iab}(\bX) \bq^i \ba^a \ba^b
+ \frac{1}{3!} H^{(0)}_{abc}(\bX) \ba^a \ba^b \ba^c
\right].
\label{actionSCSMLAflux}
\end{align}
If and only if the conditions in Theorem \ref{SCSMLAflux} are satisfied,
the action functional \eqref{actionSCSMLAflux} satisfies the classical master 
equation $\sbv{S}{S}=0$.

\subsection{SCSMs with boundaries}\label{sec:SCSMLAboundary}
\noindent
We consider a generalization to the GCSM with boundaries.
Suppose that $\Sigma$ has boundaries $\partial \Sigma \neq \emptyset$.

We have extra freedom to add boundary terms to the action functional
and can discuss consistency of boundary conditions 
with homological conditions.

In this section we take the action functional of the standard Courant
sigma model with fluxes and Lie algebroid gauging \eqref{actionSCSMLAflux}
as a bulk theory on $\Sigma$. We can consider the other CSM and gauging by 
other algebroids cases. Other cases are discussed in Appendix
\ref{sec:SCSMCAflux}, \ref{sec:GCSMLAflux} and \ref{sec:GCSMCAflux}.

Since boundaries $\partial \Sigma$ is in two dimensions,
boundary terms are integrations of degree two functions 
of superfields on $T[1]\Sigma$. We consider the following boundary terms,
\begin{align}
S_b &=\int_{T[1]\partial \Sigma} \rd^2 \sigma \rd^2 \theta
\left[
\frac{1}{2} \mu^{(2)}_{ij}(\bX) \bq^i \bq^j 
+ \mu^{(1)}_{ia}(\bX) \bq^i \ba^a 
+ \frac{1}{2} \mu^{(0)}_{ab}(\bX) \ba^a \ba^b
\right],
\label{boundaryterms}
\end{align}
where 
$\mu^{(2)}$ is a pullback of a two-forms, $\mu^{(2)} \in \Omega^2(M)$,
$\mu^{(1)}$ is a one-forms, $\mu^{(1)}$ taking a value in the pullback of 
$A^*$,
and $\mu^{(0)}$ is a zero-forms, taking a value in pullback of $\wedge^2 A^*$.

Though we can consider more general terms including $p_i, w_a$,
The choice \eqref{boundaryterms} is interesting since 
this choice has given important geometric structures 
including Dirac structures for ordinary CSMs \cite{Ikeda:2012pv},
and generalization of boundary conditions
in gauged Poisson sigma models \cite{Ikeda:2023pdr}.
In the case of the CSM, boundary geometry is described by 
Dirac structures such as the twisted Poisson structure 
\cite{Klimcik:2001vg, Park2000au}.
For the case of gauged Poisson sigma models,
boundary terms are described by momentum maps or momentum sections on 
Poisson manifolds \cite{Ikeda:2023pdr}.

Let $S_T = S_S^{\nabla} + S_A + S_F$ be the total action functional, where
$S_S + S_A$ is in Eq.~\eqref{actionSCSMLAflux}
and $S_F$ is Eq.~\eqref{SCSMLAfluxaction}.

If $\Sigma$ has boundaries, $\sbv{S}{S}$ for the bulk action $S$
gives rise to boundary terms.
Especially, $\sbv{S}{S}$ does not vanish. 
Concrete calculations give the following boundary integrations,
\begin{align}
\sbv{S}{S} &= 2 \int_{T[1]\Sigma} \rd^3 \sigma \rd^3 \theta \
\rd \left[- \bz_i^{\nabla}  \rd \bX^i 
+ \bp_i \rd \bq^i 
+ \bw_a D \ba^a
+ \bz_i^{\nabla} \bq^i
+ \rho^i_a(\bX) \bz_i^{\nabla} \ba^a 
\right.
\nonumber \\
& 
\left.
+ \frac{1}{3!} H_{ijk} \bq^i \bq^j \bq^k 
- \frac{1}{2} T_{ab}^c \ba^a \ba^b \bw_c
\right.
\nonumber \\
& 
\left.
+ \frac{1}{2} H^{(2)}_{ija} \bq^i \bq^j \ba^a 
+ \frac{1}{2} H^{(1)}_{iab} \bq^i \ba^a \ba^b
+ \frac{1}{3!} H^{(0)}_{abc} \ba^a \ba^b \ba^c
\right]
\nonumber \\ 
&= 2 \int_{T[1]\partial \Sigma} \rd^2 \sigma \rd^2 \theta
\left[- \bz_i  \rd \bX^i 
+ \bp_i \rd \bq^i 
+ \bw_a \rd \ba^a
+ \bz_i  \bq^i
+ \omega^a_{bi} \ba^b \bw_a \bq^i
\right.
\nonumber \\
& 
\left.
+ \rho^i_a(\bX) \bz_i \ba^a 
+ \frac{1}{3!} H_{ijk} \bq^i \bq^j \bq^k 
+ \frac{1}{2} C_{ab}^c \ba^a \ba^b \bw_c
\right.
\nonumber \\
& 
\left.
+ \frac{1}{2} H^{(2)}_{ija} \bq^i \bq^j \ba^a 
+ \frac{1}{2} H^{(1)}_{iab} \bq^i \ba^a \ba^b
+ \frac{1}{3!} H^{(0)}_{abc} \ba^a \ba^b \ba^c
\right],
\end{align}
Here total derivative terms become integrations on boundaries using the Stokes' theorem.

Let $S_T = S + S_b$ be the action functional with boundaries.
In order to compute $\sbv{S_T}{S_T}$,
we use formulas $\sbv{S}{\Phi}$ for every fundamental superfield $\Phi$. 
They are
\begin{align}
\delta \bX^i &= \sbv{S}{\bX^i} = \rd \bX^i - \bq^i - \rho^i_a \ba^a,
\label{bvofx}
 \\
\delta \bq^i &= \sbv{S}{\bq^i} = \rd \bq^i,
\label{bvofq} \\
\delta \ba^a &= \sbv{S}{\ba^a} = - \omega^a_{bi} \ba^b \bq^i + 
\rd \ba^a + \frac{1}{2} C^a_{bc} \ba^b \ba^c.
\label{bvofa}
\end{align}
$\sbv{S}{S_b}$ is computed using Eq.~\eqref{bvofx}--\eqref{bvofa} as
\begin{align}
\delta S_b &= \sbv{S}{S_b}
\nonumber \\ 
&= \int_{T[1]\partial \Sigma} \rd^2 \sigma \rd^2 \theta
\left[
- \frac{1}{3!} \partial_{k} \mu^{(2)}_{ij} \bq^i \bq^j \bq^k
- \frac{1}{2} (\rho^k_a \partial_k \mu^{(2)}_{ij}
+ \nabla_{i} \mu_{ja}^{(1)}) \bq^i \bq^j \ba^a
\right.
\nonumber \\ & \quad
\left.
- \left\{\left(\rho^j_a \partial_j \mu_{ib}^{(1)} 
- \frac{1}{2} C_{ab}^c \mu_{ic}^{(1)} \right) 
+ \frac{1}{2} \nabla_i  \mu^{(0)}_{ab} \right\} 
\bq^i \ba^a \ba^b
- \frac{1}{3!} \left(\rho^i_{a} \partial_i \mu^{(0)}_{bc} 
+ C_{ab}^d \mu^{(0)}_{dc} \right) \ba^a \ba^b \ba^c
\right],
\label{}
\end{align}
and $\sbv{S_F}{S_F} = 0$.

Consistency requires that boundary conditions are imposed so that 
the classical master equation $\sbv{S_T}{S_T}=0$ is satisfied 
including boundaries.
In BV-AKSZ formalism,
boundary conditions for half of superfields are imposed as input data.
Boundary conditions of another half superfields are fixed from consistency 
with equations of motion.
In the QP-manifold language, boundary conditions correspond to 
choice of a Lagrangian submanifold of the given QP-manifold.

We take the following boundary conditions,
$\bz_i = 0$,
$\bp_i =0$ and $\bw_a =0$ on boundaries,
which give interesting geometric structures on the target space 
$T^*[2](T^*[1]M \otimes A[1])$, or the corresponding non graded 
ordinary vector bundle $TM \oplus T^*M \oplus A$.

$\sbv{S_T}{S_T}=0$ is satisfied if and only if 
in addition to Eqs.~\eqref{homological01}--\eqref{homological04},
the following conditions hold,
\begin{align}
& \partial_{[k} \mu^{(2)}_{ij]} - H_{ijk} = 0,
\label{homological05}
\\
& {}^A \rd_a \mu^{(2)}_{ij} + \nabla_{[i} \mu_{j]a}^{(1)} - H^{(2)}_{ija} = 0,
\label{homological06}
\\
& {}^A \rd_{[b} \mu_{i|a]}^{(1)} + \nabla_{[i} \mu_{ab]}^{(0)} 
- H^{(1)}_{iab} = 0,
\label{homological07}
\\
& {}^A \rd_{[a} \mu_{bc]}^{(0)} 
- H^{(0)}_{abc} = 0.
\label{homological08}
\end{align}
Rewriting Eqs.~\eqref{homological05}--\eqref{homological08}
to coordinate independent forms, we obtain the theorem.
\begin{theorem}
$S_T$ is homological, i.e., satisfies the classical master equation if and only if
in addition to conditions in Theorem \ref{SCSMLAflux},
the following conditions are satisfied,
\begin{align}
& \rd \mu^{(2)}  - H = 0,
\label{hmm01}
\\
& {}^A \rd^{\nabla} \mu^{(2)} + \nabla \mu^{(1)} - H^{(2)} = 0,
\label{hmm02}
\\
& {}^A \rd^{\nabla} \mu^{(1)} + \nabla \mu^{(0)} - H^{(1)} = 0,
\label{hmm03}
\\
& {}^A \rd^{\nabla} \mu^{(0)} - H^{(0)} = 0.
\label{hmm04}
\end{align}
\end{theorem}
There is a geometric interpretation of Eqs.~\eqref{hmm01}--\eqref{hmm04}.
$\mu^{(0)}$, $\mu^{(1)}$ and $\mu^{(2)}$ 
satisfying \eqref{hmm01}--\eqref{hmm04}
are a generalization of homotopy momentum sections on 
a (pre-)$2$-plectic manifold.
In fact under Eq.~\eqref{fluxeq01}
$\mu^{(0)}$, $\mu^{(1)}$ and $\mu^{(2)}$ satisfies the definition of 
homotopy momentum sections if 
$H^{(2)} = \iota_{\rho} H$, 
$H^{(1)} = \iota_{\rho} \iota_{\rho} H$, 
and $H^{(0)} = \iota_{\rho} \iota_{\rho} \iota_{\rho} H$
\cite{Hirota:2021isx}.
Moreover $\mu^{(0)}$, $\mu^{(1)}$ and $\mu^{(2)}$ reduces
a homotopy moment map on a (pre-)multisympletic manifold
with a Lie group action \cite{Callies:2013jbu} if
a Lie algebroid $A$ is an action Lie algebroid.
If a Lie algebroid is $A = M \times \mathfrak{g}$,
the $A$-differential ${}^A \rd$ reduces to 
the Chevalley-Eilenberg differential 
of the Lie algebra $d_{CE}$ plus $\calL_\rho$, 
and under some natural assumptions, 
Eqs.~\eqref{hmm01}--\eqref{hmm04} reduces to the definition of 
a homotopy moment map \cite{Hirota:2021isx}.
Eqs.~\eqref{hmm01}--\eqref{hmm04} are regarded as a generalization to
the Courant algebroid setting.

\subsection*{Acknowledgments}
\noindent
The author would like to thank Athanasios Chatzistavrakidis 
and Miquel Cueca Ten for useful comments and discussions.

This work was supported by JSPS Grants-in-Aid for Scientific Research Number 22K03323 and by the research project, 'Higher structures in geometry and mathematical physics' in the Research Institute for Mathematical Sciences, an International Joint Usage/Research Center in Kyoto University.

\appendix

\section{Lie algebroids}\label{app:Liealgebroid}
In Appendix, we summarize definition and formulas of Lie algebroids, 
Courant algebroids and graded manifolds.

\subsection{Definitions and examples}

\begin{definition}
Let $A$ be a vector bundle over a smooth manifold $M$.
A Lie algebroid $(A, [-,-], \rhoa)$ is a vector bundle $A$ with
a bundle map $\rhoa = \rho: A \rightarrow TM$ called the anchor map, 
and a Lie bracket
$[-,-]: \Gamma(A) \times \Gamma(A) \rightarrow \Gamma(A)$
satisfying the Leibniz rule,
\begin{eqnarray}
[e_1, fe_2] &=& f [e_1, e_2] + \rhoa(e_1) f \cdot e_2,
\end{eqnarray}
where $e_1, e_2 \in \Gamma(A)$ and $f \in C^{\infty}(M)$.
\end{definition}
A Lie algebroid generalizes both a Lie algebra and the space of vector fields on a smooth manifold.
\begin{example}[Lie algebras]
Let a manifold $M$ be one point $M = \{pt \}$. 
Then a Lie algebroid is a Lie algebra $\mathfrak{g}$.
\end{example}
\begin{example}[Tangent Lie algebroids]\label{tangentLA}
If a vector bundle $A$ is a tangent bundle $TM$ and $\rhoa = \mathrm{id}$, 
then a bracket $[-,-]$ is a normal Lie bracket 
on the space of vector fields $\mathfrak{X}(M)$
and $(TM, [-,-], \mathrm{id})$ is a Lie algebroid.
It is called a \textit{tangent Lie algebroid}.
\end{example}

\begin{example}[Action Lie algebroids]\label{actionLA}
Assume a smooth action of a Lie group $G$ to a smooth manifold $M$, 
{$M \times G \rightarrow M$.}
The differential map of this action induces an infinitesimal action of the Lie algebra $\mathfrak{g}$ of $G$ on $M$.
Since $\mathfrak{g}$ acts as a differential operator on $M$,
the differential map
is a bundle map $\rho: M \times \mathfrak{g} \rightarrow TM$.
Consistency of a Lie bracket requires that $\rho$ is 
a Lie algebra morphism such that
\begin{eqnarray}
~[\rho(e_1), \rho(e_2)] &=& \rho([e_1, e_2]),
\label{almostLA}
\end{eqnarray}
where the bracket on the left-hand side 
is the standard Lie bracket of vector fields.  
These data define a Lie algebroid $(A= M \times \mathfrak{g}, [-,-], \rho)$.
known as an \textit{action Lie algebroid}.
\end{example}

\begin{example}[Poisson Lie algebroids]\label{Poisson}
%
A bivector field $\pi \in \Gamma(\wedge^2 TM)$ is called a Poisson bivector field if $[\pi, \pi]_S =0$, where $[-,-]_S$ denotes the Schouten bracket on the space of multivector fields, $\Gamma(\wedge^{\bullet} TM)$.
A smooth manifold $M$ equipped with a Poisson bivector field $\pi$ is 
called a Poisson manifold $(M, \pi)$.

Let $(M, \pi)$ be a Poisson manifold. Then, a Lie algebroid structure is induced on $T^*M$.
The map $\pi^{\sharp}$ is defined by
$\pi^{\sharp}: T^*M \rightarrow TM$ by $\bracket{\pi^{\sharp}(\alpha)}{\beta}
= \pi(\alpha, \beta)$ for all $\beta \in \Omega^1(M)$.
The anchor map is given by $\rho= - \pi^{\sharp}$,
and a Lie bracket on $\Omega^1(M)$ is defined by the Koszul bracket:
\begin{eqnarray}
[\alpha, \beta]_{\pi} = \calL_{\pi^{\sharp} (\alpha)}\beta - \calL_{\pi^{\sharp} (\beta)} \alpha - \rd(\pi(\alpha, \beta)),
\label{Koszulbracket}
\end{eqnarray}
where $\alpha, \beta \in \Omega^1(M)$.
Thus, $(T^*M, [-, -]_{\pi}, -\pi^{\sharp})$ is a Lie algebroid.
\end{example}

For a comprehensive review of Lie algebroids and their properties, see, 
for example, \cite{Mackenzie}.

For a Lie algebroid $A$, sections of the exterior algebra of $A^*$ are called \textit{$A$-differential forms}.
A differential ${}^A \rd: \Gamma(\wedge^m A^*)
\rightarrow \Gamma(\wedge^{m+1} A^*)$ on the spaces of $A$-differential forms, $\Gamma(\wedge^{\bullet} A^*)$,
called the \textit{Lie algebroid differential}, 
or the \textit{$A$-differential}, is defined as follows. 
\begin{definition}
The Lie algebroid differential ${}^A \rd: \Gamma(\wedge^m A^*)
\rightarrow \Gamma(\wedge^{m+1} A^*)$ is given by
\begin{eqnarray}
{}^A \rd \eta(e_1, \ldots, e_{m+1}) 
&=& \sum_{i=1}^{m+1} (-1)^{i-1} \rho(e_i) \eta(e_1, \ldots, 
\check{e_i}, \ldots, e_{m+1})
\nonumber \\ && 
+ \sum_{1 \leq i < j \leq m+1} (-1)^{i+j} \eta([e_i, e_j], e_1, \ldots, \check{e_i}, \ldots, \check{e_j}, \ldots, e_{m+1}),
\label{LAdifferential}
\end{eqnarray}
where $\eta \in \Gamma(\wedge^m A^*)$ and $e_i \in \Gamma(A)$.
\end{definition}
The $A$-differential satisfies $({}^A \rd)^2=0$.
It generalizes both the de Rham differential on $T^*M$ and the Chevalley-Eilenberg differential on a Lie algebra.

\subsection{Connections on Lie algebroids}\label{connectionLA}
Two types of connections are defined on Lie algebroids.
We start by introducing an ordinary connection on a vector bundle $A$.
A connection is an $\bR$-linear map,
$\nabla:\Gamma(A)\rightarrow \Gamma(A \otimes T^*M)$,
satisfying the Leibniz rule,
\begin{eqnarray}
\nabla (f e) = f \nabla e + (\rd f) \otimes e,
\end{eqnarray}
for every $e \in \Gamma(A)$ and $f \in C^{\infty}(M)$.
The dual connection on $A^*$ is defined by the equation,
\begin{eqnarray}
\rd \inner{\mu}{e} = \inner{\nabla \mu}{e} + \inner{\mu}{\nabla e},
\end{eqnarray}
for all sections $\mu \in \Gamma(A^*)$ and $e \in \Gamma(A)$,
where $\inner{-}{-}$ denotes the pairing between $A$ and $A^*$.
For simplicity, we use the same notation $\nabla$ for the dual connection.

On a Lie algebroid, we define another derivation called an $A$-connection.
Let $E$ be a vector bundle over the same base manifold $M$.
An \textit{$A$-connection} on $E$ 
with respect to the Lie algebroid $A$ is an $\bR$-linear map,
${}^A \nabla: \Gamma(E) \rightarrow \Gamma(E \otimes A^*)$,
satisfying 
\begin{eqnarray}
{}^A \nabla_e (f s) = f {}^A \nabla_e s + (\rho(e) f) s,
\end{eqnarray}
for $e \in \Gamma(A)$, $s \in \Gamma(E)$ and $f \in C^{\infty}(M)$.
%
The ordinary connection on $TM$ is a special case of an 
$A$-connection for $A=TM$, the ordinary connection 
$\nabla$ corresponds to the $TM$-connection ${}^{TM} \nabla$
on $A=TM$.

If an ordinary connection $\nabla$ is given,
an $A$-connection on $E'=A$,
$\banabla: \Gamma(A) \rightarrow \Gamma(A^* \otimes A)$ is defined by
\begin{eqnarray}
&& \banabla_{e} e^{\prime} := \nabla_{\rho(e)} {e^{\prime}},
\label{stAconnection}
\end{eqnarray}
for $e, e^{\prime} \in \Gamma(A)$.

If an ordinary vector bundle connection $\nabla$ on $A$ is given,
an important $A$-connection called the \textit{basic $A$-connection} is 
defined.
The basic $A$-connection on $A'=A$,
$\anablab: \Gamma(A) \rightarrow \Gamma(A \otimes A^*)$ 
is defined by
\begin{eqnarray}
\anablab_{e} e^{\prime} &:=& 
\nabla_{\rho(e)} {e^{\prime}} + \courante{e}{e^{\prime}},
\label{basicAconnection2}
\end{eqnarray}
for $e, e^{\prime} \in \Gamma(A)$.
A basic $A$-connection on the tangent bundle $E=TM$,
${}^A \nabla: \Gamma(TM) \rightarrow \Gamma(TM \otimes A^*)$,
is defined by
\begin{eqnarray}
\anablab_{e} v &:=& \calL_{\rho(e)} v + \rho(\nabla_v e)
= [\rho(e), v] + \rho(\nabla_v e),
\label{stAconnection1}
\end{eqnarray}
where 
$e \in \Gamma(A)$ and $v \in \mathfrak{X}(M)$.
For a $1$-form $\alpha \in \Omega^1(M)$, the \textit{basic $A$-connection} 
is given by
\begin{eqnarray}
\anablab_{e} \alpha &:=& \calL_{\rho(e)} \alpha 
+ \inner{\rho(\nabla e)}{\alpha},
\label{Econoneform}
\end{eqnarray}
as the dual connection. 
For a general discussion on connections on Lie algebroids, see
\cite{DufourZung, AbadCrainic, CrainicFernandes}.

Given a connection $\nabla$, the covariant derivative extends naturally  
to the derivation on the space of differential forms taking a value in
$\wedge^m A^*$,
$\Omega^k(M, \wedge^m A^*)$, which is called 
the \textit{exterior covariant derivative}.

Similarly, an $A$-connection ${}^A \nabla$ extends to a derivation
satisfying the Leibniz rule on the space 
$\Gamma(\wedge^m A^* \otimes \wedge^k E)$.
This extension is called \textit{the $A$-exterior covariant derivative}
${}^A \rd^{\nabla}: 
{}^A \nabla: \Gamma(\wedge^m A^* \otimes \wedge^k E)
\rightarrow \Gamma(\wedge^{m+1} A^* \otimes \wedge^k E)$
and is denoted by the same notation $\nabla$ and ${}^A\nabla$.
In this paper, we consider the $A$-exterior covariant derivatives 
on $E = T^*M$ and $E = TM$.
For $E = T^*M$, the concrete definition is as follows.
\begin{definition}
For $\Omega^k(M, \wedge^m A) = \Gamma(\wedge^m A^* \otimes \wedge^k T^*M)$,
the \textit{$A$-exterior covariant derivative}
${}^A \nabla: \Omega^k(M, \wedge^m A^*) \rightarrow \Omega^k(M, \wedge^{m+1} A^*)$ is defined by
\begin{eqnarray}
({}^A \rd^{\nabla} \alpha)(e_1, \ldots, e_{m+1})
&:=& \sum_{i=1}^{m+1} (-1)^{i-1} 
{}^A \nabla_{e_i}
(\alpha(e_1, \ldots, \check{e_i}, \ldots, e_{m+1}))
\nonumber \\ && 
+ \sum_{1 \leq i < j \leq m+1} (-1)^{i+j} \alpha([e_i, e_j], e_1, \ldots, \check{e_i}, \ldots, \check{e_j}, \ldots, e_{m+1}),
\label{Ediffconnection}
\end{eqnarray}
for $\alpha \in \Omega^k(M, \wedge^m A^*)$ and $e_i \in \Gamma(A)$.
\end{definition}
For $E=TM$, the $A$-exterior covariant derivative is given by
taking the pairing,
\begin{eqnarray}
{}^A \rd \inner{\phi}{\alpha_1, \ldots, \alpha_k} = 
\inner{{}^A \rd^{\nabla} \phi}{\alpha_1, \ldots, \alpha_k}
+ \sum_{i=1}^k (-1)^{i-1} \inner{\phi}{\alpha_1, \ldots, {}^A \rd^{\nabla} \alpha_i, \ldots, \alpha_k},
\end{eqnarray}
for $\phi \in \mathfrak{X}^k(M, \wedge^m A^*)$
and $\alpha_i \in \Omega^1(M)$.

For two connections, $\nabla$ and ${}^A \nabla$, 
we introduce tensors:
the \textit{curvature}, $R \in \Omega^2(M, A \otimes A^*)$, 
the \textit{$A$-torsion}, $T \in \Gamma(A \otimes \wedge^2 A^*)$, 
and the \textit{basic curvature} \cite{Blaom}.
$\baS \in \Omega^1(M, \wedge^2 A^* \otimes A)$.
These are defined as follows:
\beqa
R(v, v^{\prime}) &:=& [\nabla_v, \nabla_{v^{\prime}}] - \nabla_{[v, v^{\prime}]}, 
\label{curv}
\\
T(e, e^{\prime}) &:=& \nabla_{\rhoa(e)} e^{\prime} - \nabla_{\rhoa(e^{\prime})} e
- [e, e^{\prime}],
\label{Etorsion}
\\
\baS(e, e^{\prime}) &:=& 
[e, \nabla e^{\prime}] - [e^{\prime}, \nabla e]
- \nabla[e, e^{\prime}] 
- \nabla_{\rhoa(\nabla e)} e^{\prime} + \nabla_{\rhoa(\nabla e^{\prime})} e
\nonumber \\ 
&=& (\nabla T + 2 \mathrm{Alt} \, \iota_\rho R)(e, e^{\prime}),
\label{bcurv}
\eeqa
for $v, v^{\prime} \in \mathfrak{X}(M)$ and $e, e^{\prime} \in \Gamma(A)$.
for $e, e^{\prime} \in \Gamma(A)$.
Note that $\iota_\rho R \in \Omega^1(M, A^* \otimes A^* \otimes A)$ since
$\rho \in \Gamma(A^* \otimes TM)$ and $\iota_\rho$ gives the contraction 
between $TM$ and $T^*M$. Notation $\mathrm{Alt}$ means skew symmetrization on $A^* \otimes A^*$, which gives an element in $\Omega^1(M, \wedge^2 A^* \otimes A)$

\subsection{Local coordinate expressions}
In this section, we list formulas in local coordinates in Lie algebroids .

For a Lie algebroid $A$ over $M$,
$x^i$ is a local coordinate on $M$, $e_a \in \Gamma(A)$ is a basis of sections of $A$ and $e^a \in \Gamma(A^*)$ is a dual basis of sections of $A^*$. 
$i,j$, etc. are indices on $M$ and $a,b$, etc. are indices on the fiber of $A$.
Local coordinate expressions of the anchor map and the Lie bracket are
$\rho(e_a) f = \rho^i_a(x) \partial_i f$ and
$[e_a, e_b ] = C_{ab}^c(x) e_c$, where $f \in C^{\infty}(M)$ and $\partial_i = \tfrac{\partial}{\partial x^i}$.
Then, identities of $\rho^i_a$ and $C_{ab}^c$ induced from the Lie algebroid condition are
\beqa 
&& \rho_a^j \partial_j \rho_{b}^i - \rho_b^j \partial_j \rho_{a}^i = C_{ab}^c \rho_c^i,
\label{LAidentity1}
\\
&& C_{ad}^e C_{bc}^d + \rho_a^i \partial_i C_{bc}^e + \mbox{Cycl}(abc) = 0.
\label{LAidentity2}
\eeqa

Let $\omega = \omega^b_{ai} \rd x^i \otimes e^a \otimes e_b$ be 
a connection $1$-form for a connection on $A$, 
$\nabla:\Gamma(A) \rightarrow \Gamma(A \otimes T^*M)$.
For the basis vectors, the covariant derivatives are
${\nabla} e_a = - \omega_{ai}^b \rd x^i \otimes e_b$
and ${\nabla} e^a = \omega_{bi}^a \rd x^i \otimes e^b$.
Local coordinate expressions of 
covariant derivatives and the standard $A$-covariant derivatives 
on $TM$ and $A$, 
the basic $A$-connections are
\begin{eqnarray}
\nabla_i \alpha^a &=& \partial_i \alpha^a {+} \omega_{bi}^a \alpha^b,
\\
\nabla_i \beta_a &=& \partial_i \beta_a {-} \omega_{ai}^b \beta_b,
\\
\anablab_a v^i
&=& \rho_a^j \partial_j v^i - \partial_j \rho^i_a v^j
+ \rho^i_b \omega^b_{aj} v^j
= \rho_a^j \partial_j v^i - \nabla_j \rho^i_a v^j,
\\
\anablab_a \alpha_i
&=&  \rho_a^j \partial_j \alpha_i + \partial_i \rho^j_a \alpha_j
{-} \rho^j_b \omega^b_{ai} \alpha_j = 
\rho_a^j \partial_j \alpha_i + \nabla_i \rho^j_a \alpha_j,
\\
\anablab_a \alpha^b
&=& \rho^i_a \partial_i \alpha^b + \rho^i_c \omega_{ai}^b \alpha^c
+ C_{ac}^b \alpha^c
\nonumber \\
&=& \rho^i_a \nabla_i \alpha^b - T_{ac}^b \alpha^c,
\\
\anablab_a \beta_{b}
&=& \rho^i_a \partial_i \beta_b
- \rho^i_b \omega_{ai}^c \beta_c
- C_{ab}^c \beta_c
\nonumber \\
&=& \rho^i_a \nabla_i \beta_b + T_{ab}^c \beta_c,
\\
{}^A \rd^{\nabla}_{[a} \beta_{b]}
&=& \rho^i_a \partial_i \beta_b
- \rho^i_b \partial_i \beta_a 
- C_{ab}^c \beta_c 
\nonumber \\
&=& \rho^i_a(\partial_i \beta_b - \omega_{bi}^c \beta_c) 
- \rho^i_b(\partial_i \beta_a - \omega_{ai}^c \beta_c) 
+ T_{ab}^c \beta_c
\nonumber \\
&=& \rho^i_a \nabla_i \beta_b
- \rho^i_b \nabla_i \beta_a 
+ T_{ab}^c \beta_c,
\end{eqnarray}

Local coordinate expressions of 
a curvature, $R \in \Omega^2(M, A \otimes A^*)$,
an $A$-torsion, $T \in \Gamma(A \otimes \wedge^2 A^*)$, 
and a basic curvature, and $S \in \Omega^1(M, \wedge^2 A^* \otimes A)$, 
are given by
\beqa 
R_{ijb}^a &\equiv& 
\partial_i \omega_{bj}^a - \partial_j \omega_{bi}^a 
- \omega_{bi}^c \omega_{cj}^a + \omega_{bj}^c \omega_{ci}^a,
\label{curvatureona}
\\
T_{ab}^c &\equiv& 
- C_{ab}^c + \rho_a^i \omega_{bi}^c - \rho_b^i \omega_{ai}^c,
\label{atorsionona}
\\
S_{iab}^{c} &\equiv& 
\nabla_i T_{ab}^c + \rho_b^j R_{ija}^c - \rho_a^j R_{ijb}^c,
 \nonumber \\
&=& - \partial_i C^c_{ab} + \omega_{di}^c C_{ab}^d - \omega_{ai}^d C_{db}^c - \omega_{bi}^d C_{ad}^c
+ \rho_a^j \partial_j \omega_{bi}^c
- \rho_b^j \partial_j \omega_{ai}^c
\nonumber \\ && 
+ \partial_i \rho_a^j \omega_{bj}^c
- \partial_i \rho_b^j \omega_{aj}^c
+ \omega_{ai}^d \rho_d^j \omega_{bj} ^c
- \omega_{bi}^d \rho_d^j \omega_{aj} ^c
\label{bcurvatureona}
\eeqa
where $\nabla_i T_{ab}^c$ is
\beqa 
\nabla_i T_{ab}^c &\equiv& 
\partial_i T_{ab}^c
+ \omega_{di}^c T_{ab}^d - \omega_{ai}^d T_{db}^c 
- \omega_{bi}^d T_{ad}^c.
\eeqa
\section{Courant algebroids}\label{app:courantalgebroid}
In this section, we summarize definition and formulas of 
Courant algebroids.

\subsection{Definitions and examples}

\begin{definition}\label{courantdefinition}\cite{LWX}
A Courant algebroid is a vector bundle $E$ over $M$,
which has a nondegenerate symmetric bilinear form
$\bracket{-}{-}$, a bilinear operation $\courant{-}{-}$ on $\Gamma(E)$
called the Dorfman bracket, and a bundle map called an anchor map,
$\rhoe = \rho: E \longrightarrow TM$, satisfying the following properties:
%
\begin{eqnarray}
&& 1.~\courant{e_1}{\courant{e_2}{e_3}} = \courant{\courant{e_1}{e_2}}{e_3} + \courant{e_2}{\courant{e_1}{e_3}}, 
  \label{courantdef1}
\\
&& 2.~\rho(\courant{e_1}{e_2}) = [\rho(e_1), \rho(e_2)], 
  \label{courantdef2}
\\
&& 3.~\courant{e_1}{f e_2} = f \courant{e_1}{e_2}
+ (\rho(e_1)f)e_2, 
  \label{courantdef3}
 \\
&& 4.~\courant{e}{e} = \frac{1}{2} {\cal D} \bracket{e}{e},
  \label{courantdef4}
\\ 
&& 5.~\rho(e_1) \bracket{e_2}{e_3}
= \bracket{\courant{e_1}{e_2}}{e_3} + \bracket{e_2}{\courant{e_1}{e_3}},
  \label{courantdef5}
\end{eqnarray}
where 
$e, e_1, e_2, e_3 \in \Gamma(E)$, $f \in C^{\infty}(M)$ and 
${\cal D}$ is a map from $C^{\infty}(M)$ to $\Gamma (E)$, 
defined as 
$\bracket{{\cal D}f}{e} = \rho(e) f$.
\end{definition}
A Courant algebroid is encoded in the quadruple $(E, \bracket{-}{-}, \courant{-}{-}, \rho)$.

\begin{example}\label{standardCA}
The \textsl{standard Courant algebroid} with $3$-form is the Courant algebroid 
as defined below on the vector bundle $E = TM \oplus T^*M$.

The three operations of the standard Courant algebroid
are defined as follows,
\begin{align}
	\bracket{u + \alpha}{v + \beta} &= \iota_u \beta + \iota_v \alpha, \notag \\
\rhott(u+\alpha) &= u, \notag \\
\courant{u + \alpha}{v + \beta} &= [u, v] + \calL_u \beta - \iota_v \rd \alpha + \iota_u \iota_v H,
\notag
\end{align}
for $u + \alpha, v + \beta \in \Gamma(TM \oplus T^*M)$, 
where $u,v$ are vector fields, $\alpha,\beta$ are $1$-forms,
and $H \in \Omega^3(M)$ is a closed $3$-form on $M$.
\end{example}

\subsection{Connections on Courant algebroids}\label{connectionCA}
For a general discussion on connections on Courant algebroids, see
\cite{Gualtieri:2007bq, Jotz, Cueca:2019fej}.
For relations of curvatures, torsions, etc. of Courant algebroids with sigma models,
refer to \cite{Chatzistavrakidis:2023lwo}.

We begin by introducing an ordinary connection on a vector bundle $E'$
A connection is an $\bR$-linear map,
$\nabla:\Gamma(E')\rightarrow \Gamma(E' \otimes T^*M)$,
satisfying the Leibniz rule,
\begin{eqnarray}
\nabla (f s) = f \nabla s + (\rd f) \otimes s,
\end{eqnarray}
for all $s \in \Gamma(E')$ and $f \in C^{\infty}(M)$.
The dual connection on $E^{'*}$ is defined by the equation,
\begin{eqnarray}
\rd \inner{\mu}{s} = \inner{\nabla \mu}{s} + \inner{\mu}{\nabla s},
\end{eqnarray}
for all sections $\mu \in \Gamma(E^{'*})$ and $s \in \Gamma(E')$,
where $\inner{-}{-}$ denotes the pairing between $E'$ and $E^{'*}$.
For simplicity, we use the same notation $\nabla$ for the dual connection.

In this paper, we always assume that the connection and the $E$-connection satisfy
\begin{eqnarray}
\rd \bracket{e_1}{e_2} &=& \bracket{\nabla e_1}{e_2} + \bracket{e_1}{\nabla e_2},
\\
\rho(e) \bracket{e_1}{e_2} &=& \bracket{{}^E\nabla_e e_1}{e_2} + \bracket{e_1}{{}^E\nabla_e e_2}.
\end{eqnarray}
In local coordinates, $k_{AB}$ is supposed satisfying equations
\begin{eqnarray}
\nabla_i k_{AB} = {}^E \nabla_A k_{BC} = 0.
\end{eqnarray}

On a Courant algebroid, we define another derivation called an $E$-connection.
Let $E$ be a Courant algebroid over a smooth manifold $M$, and let $E'$ be a vector bundle over the same base manifold $M$.
An \textit{$E$-connection} on $E'$ 
with respect to the Courant algebroid $E$ is a
map,
${}^E \nabla: \Gamma(E') \rightarrow \Gamma(E' \otimes E^*)$,
satisfying 
\begin{eqnarray}
{}^E \nabla_e (f s) &=& f {}^E \nabla_e s + (\rho(e) f) s,
\end{eqnarray}
for $e \in \Gamma(E)$, $s \in \Gamma(E^{\prime})$ and $f \in C^{\infty}(M)$
\footnote{${}^E \nabla_{f e} s = f {}^E \nabla_e s$ does not necessarily hold
for an $E$-connection on a Courant algebroid.
A proper definition of $E$-connection called a Dorfman connection 
appears in \cite{Jotz}.
An improved definition of an $E$-connection which satisfying 
$\bR$-linearity is given by a Lie $2$-algebroid description 
of a Courant algebroid \cite{CHJL}.
}.
%

If an ordinary vector bundle connection $\nabla$ on a Courant algebroid 
$E$ is given, an $E$-connection on $E'=E$,
${}^E \nabla: \Gamma(E) \rightarrow \Gamma(E \otimes E^*)$ is defined by
\begin{eqnarray}
&& {}^E \overset{\bullet}{\nabla}_{e} e^{\prime} := \nabla_{\rho(e)} {e^{\prime}},
\label{stEconnection}
\end{eqnarray}
for $e, e^{\prime} \in \Gamma(E)$.

A speical $E$-connection induced from an ordinary vector bundle connection 
is the \textit{basic $E$-connection}.
The basic $E$-connection on $E'=E$,
$\enablab: \Gamma(E) \rightarrow \Gamma(E \otimes E^*)$ 
is defined by
\begin{eqnarray}
\enablab_{e} e^{\prime} &:=& 
\nabla_{\rho(e)} {e^{\prime}} + \courante{e}{e^{\prime}},
\label{basicEconnection2}
\end{eqnarray}
The basic $E$-connection on the tangent bundle $E'=TM$,
${}^E \nabla: \Gamma(TM) \rightarrow \Gamma(TM \otimes E^*)$ 
is defined by
\begin{eqnarray}
\enablab_{e} v 
&:=& 
[\rho(e), v] + \rho(\nabla_v e),
\label{basicEconnection1}
\end{eqnarray}
where 
$e \in \Gamma(E)$ and $v \in \mathfrak{X}(M)$.
In this paper,  we always use the basic $E$-connections 
as $E$-connections. Thus we denote 
${}^E \nabla = {}^E \nabla^{bas}$ in the paper.

Given a connection $\nabla$, the covariant derivative extends naturally  
to the derivation on the space of differential forms taking a value in
$\wedge^m E^{'*}$,
$\Omega^k(M, \wedge^m E^{'*})$, which is called 
the \textit{exterior covariant derivative}.

Similar to the Lie algebroid case, 
an $E$-connection ${}^E \nabla$ extends to a derivation
satisfying the Leibniz rule on the space 
$\Gamma(\wedge^m E^* \otimes \wedge^k E')$.
This extension is called \textit{the $E$-exterior covariant derivative}
$
{}^E \nabla: \Gamma(\wedge^m E^* \otimes \wedge^k E')
\rightarrow \Gamma(\wedge^{m+1} E^* \otimes \wedge^k E')$.
In this paper, we consider the $E$-exterior covariant derivatives 
on $E' = T^*M$ and $E' = TM$.

For two connections, $\nabla$ and ${}^E \nabla$, 
we introduce important tensors:
the \textit{curvature}, $R \in \Omega^2(M, E \otimes E^*)$, 
the \textit{$E$-torsion}, or the generalized torsion, or the Gualtieri torsion, 
$T \in \Gamma(\wedge^3 E^*)$ \cite{Gualtieri:2007bq}, 
and the \textit{basic curvature} 
$\beS \in \Omega^1(M, \wedge^3 E^*)$ \cite{Chatzistavrakidis:2023otk}.
These are defined as follows:
\begin{align}
R(v_1, v_2) &:= [\nabla_{v_1}, \nabla_{v_2}] - \nabla_{[v_1, v_2]}, 
\label{curv2}
\\
T(e_1, e_2, e_3) &:=
\bracket{{}^E \nabla_{e_1} e_2 - {}^E \nabla_{e_2} e_1 
- [e_1, e_2]_E}{e_3} 
+ \frac{1}{2} (\bracket{{}^E \nabla_{e_3} e_1}{e_2} 
- \bracket{{}^E \nabla_{e_3} e_2}{e_1}),
\label{Etorsion2}
\\
{}^E S(e_1, e_2, e_3) v &:= \bracket{\nabla_v \courante{e_1}{e_2}
- \courante{\nabla_v e_1}{e_2} - \courante{e_1}{\nabla_v e_2}
- \nabla_{\enablab_{e_2} v} e_1 + \nabla_{\enablab_{e_1} v} e_2}{e_3}
\nonumber \\ &
\qquad + \frac{1}{2} 
\left(\bracket{\nabla_{\enablab_{e_3} v} e_1}{e_2} 
- \bracket{\nabla_{\enablab_{e_3} v} e_2}{e_1}\right).
\label{bcurv2}
\end{align}
for $e_1, e_2, e_3 \in \Gamma(E)$ and $v, v_1, v_2 \in \mathfrak{X}(M)$.
Here $\courante{e_1}{e_2} = \courant{e_1}{e_2} - \courant{e_2}{e_1}$ is the Courant bracket, which is obtained by skew symmetrization of the Dorfman bracket.

\subsection{Local coordinate expressions}

In this section, we list up formulas in local coordinates in Courant algebroids .
Let $x^i$ be a local coordinate on $M$, and $e_A$ be
a local coordinate of basis of the fiber of $E$, 
and $e^A$ be a local coordinate of dual basis of the fiber of $E^*$.
We denote the inner product, the anchor map and the Dorfman bracket by
$k_{AB} = \bracket{e_A}{e_B}$, $\rho(e_A) = \rho^i_A \partial_i$,
and $[e_A, e_B]_D = - k^{CD} H_{ABC} e_D$.

Local coordinate expression of identities of operations of the Courant algebroid are 
\begin{align}
& k^{AB} \rho^i_A \rho^j_B =0,
\label{CAidentity1}
\\
& \rho^j_B \partial_j \rho^i_A - \rho^j_A \partial_j \rho^i_B
+ k^{CD} \rho^i_C H_{DAB} =0,
\label{CAidentity2}
\\
& \rho^j_D \partial_j H_{ABC} - \rho^j_A \partial_j H_{BCD}
+ \rho^j_B \partial_j H_{CDA} - \rho^j_C \partial_j H_{DAB}
\nonumber \\ & \quad
+ k^{EF} H_{EAB} H_{CDF} + k^{EF} H_{EAC} H_{DBF} 
+ k^{EF} H_{EAD} H_{BCF} 
=0.
\label{CAidentity3}
\end{align}

Let $\nabla:\Gamma(E) \rightarrow \Gamma(E \otimes T^*M)$ be an ordinary 
connection
and $\omega_A^B \otimes e_B \otimes e^A = \omega_{Ai}^B dx^i \otimes e_B \otimes e^A$ be a connection 1-form of $\nabla$.
$\omega_{Ai}^B$ satisfies
\begin{eqnarray}
\omega_{ABi} &:=& \omega^C_{Ai} k_{CB} = - k_{AC} \omega^C_{Bi}.
\end{eqnarray}

Covariant derivatives with respect to $\nabla$ is
for $\mu^A e_A \in \Gamma(E)$ and $\mu_A e^A \in \Gamma(E^*)$ are
\begin{eqnarray}
\nabla_i \mu^A &=& \partial_i \mu^A + \omega^A_{Bi} \mu^B,
\\
\nabla_i \mu_A &=& \partial_i \mu_A - \omega_{Ai}^B \mu_B.
\end{eqnarray}
Local coordinate expressions of the basic $E$-connections are 
\begin{eqnarray}
\enablab_A \mu^B &=& \rho^i_A (\partial_i \mu^B + \omega_{Ci}^B \mu^C) + T_{ABC} \mu^C,
\\
{}^E \rd^{\nabla}_{[A} \mu_{B]} &=& 
\rho^i_{[A} (\partial_i \mu_{B]} - \omega_{B]i}^C \mu_C) 
+ k^{CD} T_{ABC} \mu_D,
\\
\enablab_A v^i
&=& \rho_A^j \partial_j v^i - \partial_j \rho^i_A v^j
+ \rho^i_B \omega^B_{Aj} v^j
= \rho_A^j \nabla_j v^i - \nabla_j \rho^i_A v^j,
\\
\enablab_A \alpha_i
&=& \rho_A^j \nabla_j \alpha_i + \nabla_i \rho^j_A \alpha_j,
\end{eqnarray}
We always take a connection and an $E$-connection compatible 
the fiber metric $k_{AB}$ such that
\begin{eqnarray}
\nabla_i k_{AB} = {}^E \nabla_A k_{BC} = 0.
\end{eqnarray}
Local coordinate expressions of the curvature, the $E$-torsion 
and the basic curvature are
\begin{eqnarray}
R_{ijA}^B &=& \partial_i \omega_{Aj}^B - \partial_j \omega_{Ai}^B
- \omega_{Ai}^C \omega_{Cj}^B + \omega_{Aj}^C \omega_{Ci}^B,
\label{curvlocal2}
\\
T_{ABC} &=& H_{ABC} - \frac{1}{2} (\rho^i_{A} \omega^D_{Bi} k_{DC}
+ \mathrm{Cycl}(ABC))
\nonumber \\
&=& H_{ABC} - \rho^i_{[A|} \omega^D_{|B|i} k_{D|C]}
\nonumber \\
&=& H_{ABC} - \rho^i_{A} \omega_{BCi}
- \rho^i_{B} \omega_{CAi} - \rho^i_{C} \omega_{ABi}.
\label{Etorsionlocal2}
\\
S_{iABC} &=& \nabla_i T_{ABC} + \rho^j_A R_{ijBC}
+ \rho^j_B R_{ijCA} + \rho^j_C R_{ijAB}
\nonumber \\ 
&=& \nabla_i T_{ABC} + \rho^j_A R_{ijB}^D k_{DC}
+ \rho^j_B R_{ijC}^D k_{DA} + \rho^j_C R_{ijA}^D k_{DB}
\nonumber \\ 
&=& \partial_i H_{ABC} + [- \omega_{Ai}^D H_{DBC} 
- \rho_{A}^j \partial_j \omega_{BCi} - \partial_i \rho_{A}^j \omega_{BCj} 
+ \omega_{Ai}^D \rho^j_D \omega_{BCj} 
\nonumber \\ && 
+ (ABC \ \mbox{cyclic})].
\label{bcurvlocal2}
\end{eqnarray}
where $\nabla_j$ is the covariant derivative on $\Gamma(E \otimes \wedge^2 E^*)$,
\begin{align}
\nabla_i T_{ABC} & = \partial_i T_{ABC} - \omega_{Ai}^D T_{DBC}
- \omega_{Bi}^D T_{ADC} - \omega_{Ci}^D T_{ABD},
\end{align}

Covariantized expressions of Eq.\eqref{CAidentity1}--\eqref{CAidentity3}, 
identities of a Courant algebroid are
\begin{align}
& k^{AB} \rho^i_A \rho^j_B =0,
\label{CAidentity21}
\\
& \rho^j_B \nabla_j \rho^i_A - \rho^j_A \nabla_j \rho^i_B
+ k^{CD} \rho^i_C T_{DAB} =0,
\label{CAidentity22}
\\
& \rho^j_D \nabla_j T_{ABC} - \rho^j_A \nabla_j T_{BCD}
+ \rho^j_B \nabla_j T_{CDA} - \rho^j_C \nabla_j T_{DAB}
\nonumber \\ &
+ k^{EF} T_{EAB} T_{CDF} + k^{EF} T_{EAC} T_{DBF} 
+ k^{EF} T_{EAD} T_{BCF}  - \rho^i_{[A} \rho^j_B R_{ij|CD]}=0,
\label{CAidentity23}
\end{align}
Here $R_{ijAB} := R_{ijA}^C k_{CB}$.

Then \eqref{CAidentity23} is also written as 
\begin{align}
& \rho^j_D S_{jABC} - \rho^j_A S_{jBCD}
+ \rho^j_B S_{jCDA} - \rho^j_C S_{jDAB}
\nonumber \\ &
+ k^{EF} T_{EAB} T_{CDF} + k^{EF} T_{EAC} T_{DBF} + k^{EF} T_{EAD} T_{BCF} =0,
\end{align}

\section{Q-manifolds, QP-manifolds and AKSZ sigma models}\label{app:qmanifold}
A Lie algebroid and a Courant algebroid have descriptions by graded manifolds.
In this section, graded formulations are explained.

\subsection{Q-manifolds}
\begin{definition} 
A \textit{graded manifold} $\calM$ is a locally ringed space $(M, \calO_{\calM})$ 
over an ordinary smooth manifold $M$, whose structure sheaf $\calO_{M}$ is a $\bZ$-graded commutative algebra . The grading is compatible with the supermanifold grading, that is, a variable of even degree is commutative and a variable of odd degree is anticommutative. By definition, the structure sheaf of $M$ is locally isomorphic to $C^{\infty}(U)\otimes S^{\bullet}(V)$, where $U$ is a local chart on $M$, $V$ is a graded vector space, and $S^{\bullet}(V)$ is a free graded commutative ring on $V$.
$\calO_{\calM}$ is also denoted by $C^{\infty}(\calM)$.
Grading of an element $f \in C^{\infty}(\calM)$
is called degree and denoted by $|f|$.
For two homogeneous elements $f, g \in C^{\infty}(\calM)$, the product is graded commutative, $fg = (-1)^{|f||g|}gf$.
\end{definition} 
There are some recent reviews of graded manifolds \cite{Roytenberg:2006qz, Cattaneo:2010re, Cueca:2025aej, Ikeda:2012pv} and references therein.
We concentrate on nonnegatively graded manifolds called an N-manifold.
\begin{definition}
If an N-manifold $\calM$ has a vector field $Q$ of degree $+1$ satisfying $Q^2=0$, it is called a \textit{Q-manifold}, or a \textit{differential graded (dg) manifold}.
\end{definition} 
This vector field $Q$ is called a homological vector field.

\begin{example}[Lie algebroids]\label{LAexample}
Let $A$ be a vector bundle over $M$.
$A$ is a Lie algebroid if and only if $A[1]$ is a Q-manifold with a homological vector field \cite{Vaintrob}.
For a vector bundle $A$ over a smooth manifold $M$, $A[1]$ is a grade manifold
such that degree of fiber coordinates is shifted by one.
The concrete local expression is as follows.
Let $(x^i, a^a)$ be local coordinates on $A[1]$ of degree $(0, 1)$.
Here $i$ is the index on the base manifold $M$ and $a$ is the index of the fiber.
A vector field of degree $+1$ is locally given by
\begin{align}
Q &= \rho^i_a(x) a^a \frac{\partial}{\partial x^i}
- \frac{1}{2} C_{bc}^a(x) a^b a^c \frac{\partial}{\partial a^a},
\label{homologicalQ}
\end{align}
where $\rho^i_a(x)$ and $C^a_{bc}(x)$ are local functions.
We define a bundle map $\rho: A \rightarrow TM$ and the Lie bracket 
$[-,-]: \Gamma(A) \times \Gamma(A) \rightarrow \Gamma(A)$
as
\begin{align}
\rho(e_a) &:= \rho^i_a(x) \partial_i,
\\ 
~[e_a, e_b] &:= C_{ab}^c(x) e_c.
\end{align}
for the basis of sections $e_a \in \Gamma(A)$.
Then, $Q^2=0$ is equivalent that the two operations $(\rho, [-,-])$ satisfy the definition of a Lie algebroid on $A$:
all the identities of the anchor map and the Lie bracket in the Lie algebroid $A$ are obtained from the $Q^2=0$ condition.
\end{example}

\begin{example}[Courant algebroids]
Let $E$ be a vector bundle over a smooth manifold $M$.
Consider the shifted vector bundle $E[1]$.
Take local coordinates $(x^i, \eta^A)$ on $E[1]$, 
where $x^i$ is a local coordinate on $M$ and 
$\eta^A$ is a local coordinate of the fiber of degree one.
We introduce a bilinear bracket on $E[1]$ corresponding to the inverse of 
the fiber metric on $E$
denoted by $k^{AB} = \bracket{\eta^A}{\eta^B}$.

A Q-manifold for a Courant algebroid is defined from 
a graded cotangent bundle $\calM = T^*[2]E[1]$,
Introduce local coordinates of the fiber $T^*[2]$ by $z_i$, and $p_A = k_{AB} \eta^B$, 
where we identify $E$ and $E^*$ using by the inner product $\bracket{-}{-}$.
We have coordinates $(x^i, z_i, \eta^A)$ of degree $(0, 2, 1)$.
We consider the following vector field $Q$ of degree one on $T^*[2]E[1]$,
\begin{align}
Q &= \rho^i_A(x) q^A \frac{\partial}{\partial x^i}
- \left(\rho^i_A(x) z_i + \frac{1}{2} H_{ABC} \eta^B \eta^C \right) 
k^{AB} \frac{\partial}{\partial \eta^B}
\nonumber \\ & \quad
- \left(\partial_i \rho^j_A(x) z_j \eta^A
+ \frac{1}{3!} \partial_i H_{ABC} \eta^A \eta^B \eta^C \right) 
\frac{\partial}{\partial z_i},
\label{homologicalvf}
\end{align}
where $\rho^i_A(x)$ and $H_{ABC}(x)$ are local function on $M$.
$Q$ is a homological vector field of degree one such that $Q^2=0$ if and only if 
$E$ is a Courant algebroid,
where 
$\rho(e_A) = \rho^i_A(x) \partial_i$ is the anchor map for a basis $e_A$ of $E$, 
and $H_{ABC}$ is a structure function of the Dorfman bracket 
satisfying $[e_A, e_B]_D = H_{ABC} k^{CD} e_D$.
$Q$ induces a Courant algebroid differential 
(or an $E$-differential) ${}^E\rd$ on the complex $\Gamma(\wedge^{\bullet}E^* \oplus \calS^{\bullet} T^*M)$,
where $\calS$ is a symmetric tensor product.
\end{example}

\subsection{QP-manifolds}

\begin{definition} 
If an N-manifold $\calM$ has a graded symplectic form $\gomega$ of degree $n$ 
and a vector field $Q$ of degree $+1$ satisfies $Q^2=0$, and $\calL_Q \gomega =0$,
$(\calM, \gomega, Q)$ is called a \textit{QP-manifold} of degree $n$.
\end{definition} 
For a QP-manifold $\calM$, a graded Poisson bracket on $C^\infty ({\cal M})$ 
is defined by
$$    
\{f,g\} 
= (-1)^{|f|+n} \iota_{X_f} \delta g
= (-1)^{|f|+n+1} \iota_{X_f} \iota_{X_g} \gomega,
$$
for $f, g \in C^\infty({\cal M})$,
where the Hamiltonian vector field $X_f$ is given by the equation 
$\iota_{X_f} \omega = - \delta f$. 
In a QP-manifold with $n \neq 0$, there always exists a 
Hamiltonian function $\Theta$ such that $Q = \sbv{\Theta}{-}$.
$Q^2 = 0$ is equivalent to $\sbv{\Theta}{\Theta}=0$.
$\Theta$ is also called a homological function.

\begin{example}[Lie algebroids]\label{LAQP}
Let $A$ be a vector bundle over a smooth manifold $M$.
We consider a $2$-shifted cotangent bundle $\calM= T^*[2]A[1]$.

Take local coordinates on $T^*[2]A[1]$,
$(x^i, z_i, q^a, w_a)$ of degree $(0, 2, 1, 1)$,
where $i= 1, \ldots, \dim(M)$ and $a = 1, \ldots, \mathrm{rank}(A)$.
Since $T^*[2]A[1]$ is a (graded) cotangent bundle, it 
has a canonical graded symplectic form of degree $2$ given by
\begin{align}
\gomega &= \delta x^i \wedge \delta z_i 
+ \delta a^a \wedge \delta w_a.
\label{gsymplecticl}
\end{align}

A QP-manifold structure on $T^*[2]A[1]$ is defined by introducing 
a homological function $\Theta$ of degree three. We consider 
the following function,
\begin{align}
\Theta &= \rho^i_a(x) z_i  a^a + \frac{1}{2} C_{ab}^c(x) a^a a^b w_c.
\label{CAhomologicalfnl}
\end{align}
Imposing the condition $\sbv{\Theta}{\Theta}=0$, a QP-manifold structure on 
$T^*[2]A[1]$ is given.
\begin{theorem}
A QP-manifold with the homological function \eqref{CAhomologicalfnl}
has one to one correspondence to a Lie algebroid $A$.
\end{theorem}
A Lie algebroid structure on $A$ is recovered from a 
QP-manifold $T^*[2]A[1]$ by derived brackets.
In fact, operations of a Lie algebroid are given by
\begin{align}
\sbv{e_1}{e_2} &= -\sbv{\sbv{e_1}{\Theta}}{e_2},
\\
\rho(e)f &= -\sbv{\sbv{e}{\Theta}}{f}.
\end{align}
Here $f \in C^{\infty}(M)$ and $e, e_1, e_2 \in C^{\infty}_1(A[1])$, 
where $C^{\infty}_1(A1])$ is the space of degree one functions,
which is identified to the space of sections $\Gamma(A)$.
\end{example}

\begin{example}[Courant algebroids]\label{CAQP}
Let $E$ be a vector bundle over a smooth manifold $M$.
Assume an inner product on the fiber $\bracket{-}{-}$.
The local coordinate expression of the inner product is 
$k_{AB} = \bracket{e_A}{e_B}$ for the basis $e_A \in \Gamma(E)$.
We consider a $2$-shifted cotangent bundle $\calM= T^*[2]E[1]$.

Take local coordinates on $T^*[2]E[1]$,
$(x^i, z_i, \eta^A)$ of degree $(0, 2, 1)$,
where $i= 1, \ldots, \dim(M)$ and $A= 1, \ldots, \mathrm{rank}(E)$.
Here $E$ and $E^*$ is identified by the inner product $\bracket{-}{-}$.
Since $T^*[2]E[1]$ is a (graded) cotangent bundle, it 
has a canonical graded symplectic form of degree $2$ given by
\begin{align}
\gomega &= \delta x^i \wedge \delta z_i 
+ \frac{1}{2} \delta \eta^A \wedge \delta (k_{AB} \eta^B).
\label{gsymplectic}
\end{align}

A QP-manifold structure on $T^*[2]E[1]$ is defined by introducing a homological 
function $\Theta$ of degree three. The most general form is 
\begin{align}
\Theta &= \rho^i_A(x) z_i \eta^A + \frac{1}{3!} H_{ABC}(x) \eta^A \eta^B \eta^C.
\label{CAhomologicalfn}
\end{align}
Imposing the condition $\sbv{\Theta}{\Theta}=0$, a QP-manifold structure on 
$T^*[2]E[1]$ is given.
There are a basic theorem for a QP-manifold of degree two.
\begin{theorem}\cite{Roy01}
A QP-manifold of degree two has one to one correspondence to a Courant algebroid $E$.
\end{theorem}
A Courant algebroid structure on $E$ is recovered from a QP-manifold of degree two $T^*[2]E[1]$ by derived brackets.
In fact, operations of a Courant algebroid are given by
\begin{align}
\courant{e_1}{e_2} &= -\sbv{\sbv{e_1}{\Theta}}{e_2},
\\
\rho(e)f &= -\sbv{\sbv{e}{\Theta}}{f},
\\
\calD f &= \sbv{\Theta}{f}.
\end{align}
Here $f \in C^{\infty}(M)$ and $e, e_1, e_2 \in C^{\infty}_1(E[1])$, 
where $C^{\infty}_1(E[1])$ is the space of degree one functions,
which is identified to $\Gamma(E)$.

On the other hand, for a Courant algebroid $E$,
the homological function \eqref{CAhomologicalfn}
on $T^*[2]E[1]$ is constructed from the anchor map $\rho$
and the Dorfman bracket $\courant{-}{-}$ by
taking $\rho(e_A) = \rho^i_A(x) \partial_i$, 
$\bracket{e_A}{e_B} = k_{AB}$
and $\courant{e_A}{e_B} = H_{ABC}(x) k^{CE} e_E$
for the basis $e_A \in \Gamma(E)$.
\end{example}

\begin{example}\label{standardCAQP}
As a special case of the previous example \ref{CAQP},
we consider the standard Courant algebroid in Example \ref{standardCA}.
In this case, the vector bundle is $E= TM \oplus T^*M$ and 
$k_{AB} =
\begin{pmatrix}
   0 & 1 \\
   1 & 0
\end{pmatrix}
$ for the basis $e_a \in \Gamma(E)$.
The $2$-shifted cotangent bundle is $\calM= T^*[2]T[1]M= T^*[2]T^*[1]M$.

Local coordinates on $T^*[2]T^*[1]M$ are 
$(x^i, z_i, q^i, p_i)$ of degree $(0, 2, 1, 1)$, 
where $\eta^A= (q^i, p_i)$.

A canonical graded symplectic form of degree $2$ 
on $T^*[2]T^*[1]M$ is given by
\begin{align}
\gomega &= \delta x^i \wedge \delta z_i 
+ \delta q^i \wedge \delta p_i.
\label{gsymplectic2}
\end{align}

A QP-manifold structure on $T^*[2]T^*[1]M$ is defined by introducing a homological 
function $\Theta$ of degree three. The homological function of
the standard Courant algebroid is
\begin{align}
\Theta &= z_i q^i + \frac{1}{3!} H_{ijk}(x) q^i q^j q^k,
\label{CAhomologicalfn2}
\end{align}
where $H = \frac{1}{3!} H_{ijk}(x) 
\rd x^i \wedge \rd x^j \wedge \rd x^k \in \Omega^3(M)$
is a closed $3$-form on $M$.
Imposing the condition $\sbv{\Theta}{\Theta}=0$, a QP-manifold structure on 
$T^*[2]T^*[1]M$ induces a standard Courant algebroid structure on 
$TM \oplus T^*M$.
Degree one function $u^i(x) p_i + \alpha_i(x) q^i
\in C_1^{\infty}(T^*[2]T^*[1]M)$ is identified to
a section $u + \alpha = u^i(x) \partial_i + \alpha_i(x) \rd x^i
\in \Gamma(TM \oplus T^*M)$.
As in the general Courant algebroid in Example \ref{CAQP},
derived brackets give operations of standard Courant algebroids.
\end{example}

\subsection{AKSZ sigma models}\label{sec:AKSZ}
We briefly review sigma models called AKSZ sigma models 
\cite{Alexandrov:1995kv}.
For more details, refer to articles, for instance, \cite{Cattaneo:2001ys,  Roytenberg:2006qz, Ikeda:2012pv, Ikeda:2025}

If a QP-manifold $(\calM, \gomega, \Theta)$ of degree $n$ is given, 
a QP-manifold structure on a mapping space $\Map(T[1]\Sigma, \calM)$
is defined. It is called an AKSZ sigma model.
Here $\Sigma$ is an ordinary smooth manifold.
If $\Sigma$ is in $n+1$-dimensions, the AKSZ sigma model is equivalent to 
BV-formalism of a gauge theory, 
If $\Sigma$ is in $n$-dimensions, the AKSZ sigma model is equivalent to 
BFV-formalism. 

Concrete construction is as follows.
We assume a super differential $\rd$ on $T[1]\Sigma$ induced from the de Rham differential on $\Sigma$ and a Berezin integration on $T[1]\Sigma$.
Let $\bX: \Sigma \rightarrow M$ for $\Sigma$ and the base manifold 
$M$ of $\calM$, 
Let $\vartheta$ be a Liouville $1$-form of $\gomega$ such that 
$\omega = - \delta \vartheta$.

Then the graded symplectic form on $\Map(T[1]\Sigma, \calM)$ is
given by 
\begin{align}
\ggomega &= \int_{T[1]\Sigma} \bX^* \gomega,
\end{align}
The BV-AKSZ action functional is given by
\begin{align}
S &= \int_{T[1]\Sigma} (\iota_{\rd} \bX^* \vartheta + \bX^* \Theta),
\end{align}
where $\bX:T[1]\Sigma \rightarrow M$ is a map from $T[1]\Sigma$ from 
the degree zero base manifold $M$,
and $\rd$ is the super derivative on $T[1]\Sigma$.
Local coordinates on $\calM$ is mapped to corresponding superfields.
We can prove that $S$ is a homological function, i.e., satisfies 
$\sbv{S}{S}=0$, which is the classical master equation in BV formalism.
A brief interpretation of $S$ is that it generates a 
homological vector field induced from two differentials on $T[1]\Sigma$
and $\calM$, $\rd + Q = \sbv{S}{-}$.

\subsection{Manifestly covariant expressions}\label{manicovariant}
Here properties of the grade manifold of degree two $T^*[2]E[1]$ 
are explained in local coordinates.

Let $M^B_A(x)$ be a transition function of the change of local basis 
$e_A$ to $e^{\prime}_A = M^B_A(x) e_B$ on the vector bundle $E$.
Moreover we consider arbitrary coordinate changes
$x^{\prime i} = x^{\prime i}(x)$ on the base manifold $M$.
Transformations of local coordinate on $T^*[2]E[1]$ 
from $(x^i, z_i, \eta^A)$ to $(x^{\prime i}, z^{\prime}_i, \eta^{\prime A})$ 
are given by
\begin{align}
& x^{\prime i} = x^{\prime i}(x),
\label{coordinatetransf1}
\\
& z^{\prime}_i = \frac{\partial x^j}{\partial x^{\prime i}} z_j 
+ \frac{1}{2} k_{AB} M^B_C \partial^{\prime}_i M_D^C \eta^A \eta^D,
\label{coordinatetransf2}
\\
& \eta^{\prime A} = M^A_B \eta^B,
\label{coordinatetransf3}
\end{align}
where $x^{\prime i}(x)$ is an arbitrary smooth functions of $x$.
Therefore a graded manifold $T^*[2]E[1]$ is not a cotangent bundle 
in the ordinary sense.
The graded symplectic form is invariant under local coordinate
transformations \eqref{coordinatetransf1}--\eqref{coordinatetransf3},
\begin{align}
\gomega^{\prime} &= \gomega.
\end{align}
Since $z_j$ does not transform covariantly, 
we take covariantized local coordinates as
\begin{align}
x^{\nabla i} &= x^{i},
\\
z^{\nabla}_i &
= z_i + \frac{1}{2} \omega_{AB i} \eta^A \eta^B
= z_i + \frac{1}{2} \omega_{Ai}^C k_{CB} \eta^A \eta^B
= z_i - \frac{1}{2} k_{AC} \omega_{Bi}^C \eta^A \eta^B,
\\
\eta^{\nabla A} &= \eta^A,
\end{align}
In fact, $z^{\nabla}_i $ transforms covariantly under local coordinate transformations,
 \eqref{coordinatetransf1}--\eqref{coordinatetransf3}.
Nonzero Poisson brackets for local coordinates are
\begin{align}
\sbv{x^i}{z_i^{\nabla}} &= \delta^i_j,
\\
\sbv{\eta^A}{\eta^B} &= k^{AB},
\\
\sbv{z_i^{\nabla}}{\eta^A} &= \omega_{Bi}^A \eta^B,
\\
\sbv{z_i^{\nabla}}{z_j^{\nabla}} &= 
- \frac{1}{2} R_{ijB}^C k_{CA} \eta^A \eta^B.
\end{align}

The covariantized homological function on $T^*[2]E[1]$ for a Courant algebroid is
\begin{align}
\Theta &= \rho^i_A(x) z^{\nabla}_i \eta^A 
+ \frac{1}{3!} T_{ABC}(x) \eta^A \eta^B \eta^C,
\label{hfuncov}
\end{align}
Graded Poisson brackets between the homological function \eqref{hfuncov}
with local coordinates are
\begin{align}
\{\Theta, f(x) \} &= - \rho^i_A \eta^A \partial_i f(x),
\\
\{\Theta, \eta^A \} &=  k^{AB} \rho^i_B z^{\nabla}_i
+ \rho^i_B \omega^A_{Ci} \eta^B \eta^C 
+ \frac{1}{2} k^{AD} T_{DBC} \eta^B \eta^C,
\\
\{\Theta, z^{\nabla}_i \} &= \nabla_i \rho^j_A z^{\nabla}_j \eta^A
+ \frac{1}{3!} S_{iABC} \eta^A \eta^B \eta^C.
\end{align}

\section{SCSMs with Courant algebroid gauging}
\label{sec:SCSMCAflux}
\noindent
In this section, we extend gauged Courant sigma models by adding flux terms and 
boundary terms.
In order to introduce boundary terms, assume that a three dimensional manifold 
$\Sigma$ has boundaries.
We have considered the SCSM with Lie algebroid gauging in Section \ref{sec:CSMflux}
In preceding sections, the SCSM with Courant algebroid gauging, 
and the general CSM with Lie algebroid and Courant algebroid gauging
are discussed.

\subsection{Introduction of fluxes}
In this section, the SCSM with Courant algebroid gauging is analyzed.
We can add more terms to the homological function 
$\Theta^{\nabla}= \Thetas^{\nabla} + \Thetae^{\nabla}$
in Eqw.~\ref{SCSMCAhomological01} and \ref{SCSMCAhomological02}.
We add following terms,
\begin{align}
\Thetaf &=
\frac{1}{2} H^{(2)}_{ija}(x) q^i q^j a^a 
+ \frac{1}{2} H^{(1)}_{iab}(x) q^i a^a a^b
+ \frac{1}{3!} H^{(0)}_{abc}(x) a^a a^b a^c.
\end{align}
Here $H^{(2)} = \frac{1}{2} H^{(2)}_{ija}(x) \rd x^i \wedge \rd x^j \otimes e^a
\in \Omega^2(M, E^*)$ 
is a two-form taking a value in $E^*$,
$H^{(1)} = \frac{1}{2} H^{(1)}_{iab}(x) \rd x^i \otimes e^a \wedge e^b.
\in \Omega^1(M, \wedge^2 E^*)$
is a one-form taking a value in $\wedge^2 E^*$,
and $H^{(0)}= \frac{1}{3!} H^{(0)}_{abc}(x) e^a \wedge e^b \wedge e^c.
\in \Omega^0(M, \wedge^3 E^*)$
is a zero-form taking a value in $\wedge^3 E^*$.

Poisson brackets including $\Thetaf$ 
are calculated using Poisson brackets of coordinates as follows,
\begin{align}
\sbv{\Thetas^{\nabla}}{\Thetaf} 
& = - \frac{1}{3!} \nabla_{[i} H^{(2)}_{jk]a}(x) q^i q^j q^k a^a 
+ \frac{1}{4} \nabla_{[i}H^{(1)}_{j]ab}(x) q^i q^j a^a a^b
- \frac{1}{3!} \partial_{i}H^{(0)}_{abc}(x) q^i a^a a^b a^c.
\label{TTC04}
\\
\sbv{\Thetae}{\Thetaf} 
& = 
- \frac{1}{4} (\banabla_{[a} H^{(2)}_{ij|b]}
+ m^{cd} T_{abc} H^{(2)}_{ijd} ) q^i q^j a^a a^b
\nonumber \\ & \quad
+ \frac{1}{3!} (\banabla_{[a|} H^{(1)}_{i|bc]} 
+ m^{de} T_{[ab|d} H^{(1)}_{i|e|c]} ) q^i a^a a^b a^c
\nonumber \\ & \quad
- \frac{1}{4!} (\banabla_{[a}H^{(0)}_{bcd}
+ m^{ef} T_{[ab|e} H^{(0)}_{f|cd]}) a^a a^b a^c a^d.
\nonumber \\ & \quad
+ \frac{1}{2} \rho^i_a m^{ab} H^{(2)}_{jkb} z^{\nabla}_i q^j q^k
- \rho^i_a m^{ab} H^{(1)}_{jbc} z^{\nabla}_i q^j a^c
+ \frac{1}{2} \rho^i_a m^{ab} H^{(0)}_{bcd} z^{\nabla}_i a^c a^d.
\label{TTC05}
\\
\sbv{\Thetaf}{\Thetaf}  
& = \frac{1}{4!} H_{[ij|a}^{(2)} m^{ad} H_{|kl]d}^{(2)} q^i q^j q^k q^l
- \frac{2}{3!} H_{[ij|a}^{(2)} m^{ad} H_{|k]de}^{(1)} q^i q^j q^k a^e
\nonumber \\ & \quad
+ \frac{2}{4} H_{ija}^{(2)} m^{ad} H_{def}^{(0)} q^i q^j a^e a^f
+ \frac{1}{4} H_{[i[ab|}^{(1)} m^{ad} H_{j|de]}^{(1)} q^i q^j a^b a^e
\nonumber \\ & \quad
- \frac{2}{3!} H_{ia[b|}^{(1)} m^{ad} H_{d|ef]}^{(0)} q^i a^b a^e a^f
+ \frac{1}{4!} H_{a[bc|}^{(0)} m^{ad} H_{d|ef]}^{(0)} a^b a^c a^e a^f.
\label{TTC06}
\end{align}
The total homological function is deformed to
$\Theta^{\nabla} = \Thetas^{\nabla} + \Thetae + \Thetaf$.
Combining above equations Eqs.~\eqref{TTC04}--\eqref{TTC06},
with Eqs.~\eqref{TT01}--\eqref{TT03},
The condition $\sbv{\Theta^{\nabla}}{\Theta^{\nabla}}=0$ is satisfied if 
following equations hold,
\begin{align}
& \rho^i_a m^{ab} H^{(2)}_{jkb} = 0,
\\
& \nabla_j \rho^i_a + \rho^i_b m^{bc} H^{(1)}_{jca} = 0,
\\
& \rho^i_a m^{ab} H^{(0)}_{bcd}  = 0,
\\
& H_{[ij|a}^{(2)} m^{ad} H_{|kl]d}^{(2)} =0,
\\
& {}^A \rd_{a} H_{ijk}(x) 
-  \nabla_{[i} H^{(2)}_{jk]a}(x) 
- H_{[ij|b}^{(2)} m^{bc} H_{|k]ca}^{(1)} = 0,
\\
& 2 R_{ija}^c m_{cb} 
+ \nabla_{[i}H^{(1)}_{j]ab}(x) 
- {}^A \rd_{[a} H^{(2)}_{ij|b]}(x) 
\nonumber \\ & \quad
+ H_{ijc}^{(2)} m^{cd} H_{dab}^{(0)}
+ H_{[i[ca|}^{(1)} m^{cd} H_{j|db]}^{(1)}
= 0,
\\
& - S_{iabc}
- \partial_{i}H^{(0)}_{abc}(x)
+ {}^A \rd_{[a|}H^{(1)}_{i|bc]}(x) 
- H_{id[a|}^{(1)} m^{de} H_{e|bc]}^{(0)}
= 0,
\\
& {}^A \rd_{[a}H^{(0)}_{bcd]}(x) 
+ H_{e[ab|}^{(0)} m^{ef} H_{f|cd]}^{(0)} = 0.
\end{align}
The curvature $R$ and the basic curvature ${}^A S$ do not necessarily 
vanish any more.

The corresponding term to $\Thetaf$ in the AKSZ action is
\begin{align}
S_F
&=\int_{T[1]\Sigma} \rd^3 \sigma \rd^3 \theta
\left[\frac{1}{2} H^{(2)}_{ija}(\bX) \bq^i \bq^j \ba^a 
+ \frac{1}{2} H^{(1)}_{iab}(\bX) \bq^i \ba^a \ba^b
+ \frac{1}{3!} H^{(0)}_{abc}(\bX) \ba^a \ba^b \ba^c
\right].
\end{align}
Then, the total action functional is a deformation 
of the action functional \eqref{SCSMCAaction} by $S_F$,
\begin{align}
S &= S_S^{\nabla} + S_E + S_F
\nonumber \\ 
&=\int_{T[1]\Sigma} \rd^3 \sigma \rd^3 \theta
\left[- \bz_i ^{\nabla} \rd \bX^i 
+ \bp_i \rd \bq^i 
+ \frac{1}{2} m_{ab} \ba^a D \ba^b
+ \bz_i ^{\nabla} \bq^i
+ \rho_a^i \bz_i^{\nabla} \ba^a
\right.
\nonumber \\
& 
\left.
+ \frac{1}{3!} H_{ijk}(\bX) \bq^i \bq^j \bq^k 
+ \frac{1}{2} T_{abc}(\bX) \ba^a \ba^b \ba^c
\right.
\nonumber \\
& 
\left.
+ \frac{1}{2} H^{(2)}_{ija}(\bX) \bq^i \bq^j \ba^a 
+ \frac{1}{2} H^{(1)}_{iab}(\bX) \bq^i \ba^a \ba^b
+ \frac{1}{3!} H^{(0)}_{abc}(\bX) \ba^a \ba^b \ba^c
\right].
\label{SCAMCAfluxaction}
\end{align}

\subsection{SCSMs with boundaries}
\noindent
In this section, we introduce boundary terms.
Suppose that $\Sigma$ has boundary $\partial \Sigma \neq \emptyset$.
Let $S = S_S^{\nabla} + S_A + S_F$ 
in Eq.~\eqref{SCAMCAfluxaction} be the AKSZ action on the bulk $\Sigma$.
Then the classical master equation does not hold any more, and 
$\sbv{S}{S}$ is as follows,
\begin{align}
\sbv{S}{S} 
&= 2 \int_{T[1]\Sigma} \rd^3 \sigma \rd^3 \theta \
\rd 
\left[- \bz_i^{\nabla} \rd \bX^i 
+ \bp_i \rd \bq^i 
+ \frac{1}{2} m_{ab} \ba^a D \ba^b
+ \bz_i ^{\nabla} \bq^i
+ \rho_a^i \bz_i^{\nabla} \ba^a
\right.
\nonumber \\
& 
\left.
+ \frac{1}{3!} H_{ijk} \bq^i \bq^j \bq^k 
+ \frac{1}{3!} T_{abc} \ba^a \ba^b \ba^c
\right.
\nonumber \\
& 
\left.
+ \frac{1}{2} H^{(2)}_{ija} \bq^i \bq^j \ba^a 
+ \frac{1}{2} H^{(1)}_{iab} \bq^i \ba^a \ba^b
+ \frac{1}{3!} H^{(0)}_{abc} \ba^a \ba^b \ba^c
\right]
\nonumber \\
&= 2 \int_{T[1]\Sigma} \rd^3 \sigma \rd^3 \theta \
\rd 
\left[- \bz_i \rd \bX^i 
+ \bp_i \rd \bq^i 
+ \frac{1}{2} m_{ab} \ba^a \rd \ba^b
+ \bz_i \bq^i
- \frac{1}{2} \omega^c_{ai} m_{cb} \ba^a \ba^b \bq^i
\right.
\nonumber \\
& 
\left.
+ \rho_a^i \bz_i \ba^a
+ \frac{1}{3!} H_{ijk} \bq^i \bq^j \bq^k 
+ \frac{1}{3!} C_{abc} \ba^a \ba^b \ba^c
\right.
\nonumber \\
& 
\left.
+ \frac{1}{2} H^{(2)}_{ija} \bq^i \bq^j \ba^a 
+ \frac{1}{2} H^{(1)}_{iab} \bq^i \ba^a \ba^b
+ \frac{1}{3!} H^{(0)}_{abc} \ba^a \ba^b \ba^c
\right],
\label{SCSMCAboundary01}
\end{align}
where total derivative terms become integrations on boundaries using the Stokes' theorem.
We can introduce boundary terms to describe various
boundary conditions on the theory.
Though there are possible consistent boundary terms,
we consider the following boundary terms,
\begin{align}
S_b &=\int_{T[1]\partial \Sigma} \rd^2 \sigma \rd^2 \theta
\left[\frac{1}{2} \mu^{(2)}_{ij}(\bX) \bq^i \bq^j 
+ \mu^{(1)}_{ia}(\bX) \bq^i \ba^a 
+ \frac{1}{2} \mu^{(0)}_{ab}(\bX) \ba^a \ba^b
\right].
\end{align}
Note that we can consider more general boundary terms.
The total action is $S_T= S + S_b$. In order to compute
$\sbv{S_T}{S_T}$, we use following formulas,
\begin{align}
\delta \bX^i &= \sbv{S}{\bX^i} = \rd \bX^i - \bq^i - \rho^i_a \ba^a,
\label{SPhi01}
\\
\delta \bq^i &= \sbv{S}{\bq^i} = \rd \bq^i,
\label{SPhi02}
\\
\delta \ba^a &= \sbv{S}{\ba^a} = \omega^a_{bi} \ba^b \bq^i 
+ \rd \ba^a + \frac{1}{2} m^{ad} C_{dbc} \ba^b \ba^c
\nonumber \\ & \quad 
+ \frac{1}{2} m^{ab} H^{(2)}_{ijb} \bq^i \bq^j 
- m^{ab} H^{(1)}_{ibc} \bq^i \ba^c
+ \frac{1}{2} m^{ad} H^{(0)}_{dbc} \ba^b \ba^c.
\label{SPhi03}
\end{align}
Using Eqs.~\eqref{SPhi01}--\eqref{SPhi03}, 
$\delta S_b = \sbv{S}{S_b}$ is calculated as
\begin{align}
\delta S_b 
&= \int_{T[1]\partial \Sigma} \rd^2 \sigma \rd^2 \theta
\left[- \frac{1}{3!} \left\{\partial_{[k} \mu^{(2)}_{ij]} 
+ m^{ab} H^{(2)}_{[ij|a} \mu^{(1)}_{|k]b]} \right\} \bq^i \bq^j \bq^k 
\right.
\nonumber \\ & \quad
\left.
- \frac{1}{2} \left\{ 
\rho^k_a \partial_{k} \mu^{(2)}_{ij} 
+ \nabla_{[i} \mu_{j]a}^{(1)} 
+ m^{bc} H^{(1)}_{[i|ba} \mu^{(1)}_{|j]c]} 
+ m^{bc} H^{(2)}_{ijb} \mu^{(0)}_{ca} 
\right\}\bq^j \bq^i \ba^a
\right.
\nonumber \\ & \quad
\left.
- \frac{1}{2} \left\{\left(\rho^j_{[b} \partial_j \mu_{i|a]}^{(1)} 
+ m^{cd} C_{dab} \mu_{ic}^{(1)} \right) 
+ \nabla_i \mu^{(0)}_{ab} 
+ m^{cd} H^{(0)}_{dab} \mu^{(1)}_{|i]c]} 
+ m^{cd} H^{(1)}_{ic[b|} \mu^{(0)}_{d|a]} 
\right\} 
\bq^i \ba^a \ba^b
\right.
\nonumber \\ & \quad
\left.
- \frac{1}{3!} \left(\rho^i_{[a|} \partial_i \mu^{(0)}_{|bc]} 
+ m^{de} C_{e[cb|} \mu^{(0)}_{d|a]} 
+ m^{de} H^{(0)}_{d[bc|} \mu^{(0)}_{e|a]} 
\right) \ba^a \ba^b \ba^c
\right].
\label{SCSMCAboundary02}
\end{align}
In this case, $\sbv{S_b}{S_b}$ is not zero. It is
\begin{align}
\sbv{S_b}{S_b}
&= \int_{T[1]\partial \Sigma} \rd^2 \sigma \rd^2 \theta
\left[- m^{ab} \mu^{(1)}_{ia} \mu^{(1)}_{jb} \bq^i \bq^j
+ 2 m^{ab} \mu^{(1)}_{ia} \mu^{(0)}_{bc} \bq^i \ba^c
- m^{ac} \mu^{(0)}_{ab} \mu^{(0)}_{cd} \ba^b \ba^d
\right].
\label{SCSMCAboundary03}
\end{align}
We impose similar boundary conditions as in Section 
\ref{sec:SCSMLAboundary}.
$\bz_i = 0$,
$\bp_i =0$ on boundaries.
For $\ba^a$, we impose the condition such that
half of degrees of freedom of $\ba^a$ are zero.
Concretely, the boundary condition of $\ba^a$ is written as
the inner product $m_{ab} \ba^a \rd \ba^b =0$ on boundaries.

Using Eqs.~\eqref{SCSMCAboundary01}--\eqref{SCSMCAboundary03},
the condition $\sbv{S_T}{S_T}=0$ 
imposes the following conditions to $\mu^{(0)}$, $\mu^{(1)}$ and $\mu^{(2)}$ 
in boundary terms,
\begin{align}
& \partial_{[k} \mu^{(2)}_{ij]} + m^{ab} H^{(2)}_{[ij|a} \mu^{(1)}_{|k]b]} 
- H_{ijk} = 0,
\\
& 
{}^A \rd_a \mu^{(2)}_{ij} + \nabla_{[i} \mu_{j]a}^{(1)} 
+ m^{bc} H^{(1)}_{[i|ba} \mu^{(1)}_{|j]c]} 
+ m^{bc} H^{(2)}_{ijb} \mu^{(0)}_{ca} 
- H^{(2)}_{ija} = 0,
\\
& {}^A \rd_{[b} \mu_{i|a]}^{(1)} + \nabla_{[i} \mu_{ab]}^{(0)} 
+ m^{cd} H^{(0)}_{dab} \mu^{(1)}_{|i]c]} 
+ m^{cd} H^{(1)}_{ic[b|} \mu^{(0)}_{d|a]} 
- H^{(1)}_{iab} = 0,
\\
& {}^A \rd_{[a} \mu_{bc]}^{(0)} 
+ m^{de} H^{(0)}_{d[bc|} \mu^{(0)}_{e|a]} 
- H^{(0)}_{abc} = 0,
\\
& m^{ab} \mu^{(1)}_{ia} \mu^{(1)}_{jb} =0,
\\
& m^{ab} \mu^{(1)}_{ia} \mu^{(0)}_{bc} =0,
\\
& m^{ac} \mu^{(0)}_{ab} \mu^{(0)}_{cd} =0.
\end{align}

\section{General CSMs with Lie algebroid gauging}
\label{sec:GCSMLAflux}
\noindent
We deform the model in Section \ref{sec:GCSMLA}
by adding flux terms and boundary terms.

\subsection{Introduction of fluxes}
To the homological function 
$\Theta^{\nabla} = \Thetac^{\nabla} + \Thetaa$
in Eqs.~\eqref{GCSMLAhomological01} and \eqref{GCSMLAhomological02},
we add following terms,
\begin{align}
\Theta_F &=
\frac{1}{2} H^{(2)}_{ABa}(x) \eta^A \eta^B a^a 
+ \frac{1}{2} H^{(1)}_{Aab}(x) \eta^A a^a a^b
+ \frac{1}{3!} H^{(0)}_{abc}(x) a^a a^b a^c.
\label{fluxterm03}
\end{align}
Here $H^{(2)} = \frac{1}{2} H^{(2)}_{ABa}(x) e^A \wedge e^B \otimes e^a
\in \Gamma(M, \wedge^2 E^* \otimes A^*)$ 
is an $E$-two-form taking a value in $A^*$,
$H^{(1)} = \frac{1}{2} H^{(1)}_{Aab}(x) e^A \otimes e^a \wedge e^b.
\in \Gamma(M, E^* \otimes \wedge^2 A^*)$
is an $E$-one-form taking a value in $\wedge^2 A^*$,
and $H^{(0)}= \frac{1}{3!} H^{(0)}_{abc}(x) e^a \wedge e^b \wedge e^c.
\in C^{\infty}(M, \wedge^3 A^*)$
is an $E$-zero-form taking a value in $\wedge^3 A^*$.
The extended function is 
$\Theta^{\nabla} = \Thetac^{\nabla} + \Thetaa + \Thetaf$.

Poisson brackets including $\Thetaf$ are
\begin{align}
\sbv{\Thetac^{\nabla}}{\Thetaf} 
& = - \frac{1}{3!} \rhoe^i_{[A|} \nabla_{i} H^{(2)}_{|DE]a}
\eta^A \eta^D \eta^E a^a 
+ \frac{1}{4} \rhoe^i_{[A|} \nabla_{i} H^{(1)}_{|D]ab} \eta^A \eta^B a^a a^b
\nonumber \\ & \quad
- \frac{1}{3!} \rhoe^i_{A} \partial_{i} H^{(0)}_{abc} \eta^A a^a a^b a^c.
\nonumber \\ & \quad
+ \rhoe^i_A k^{AD}H^{(2)}_{DEa} z^{\nabla}_i \eta^E a^a
+ \frac{1}{2} \rhoe^i_A k^{AD} H^{(1)}_{Dab} z^{\nabla}_i a^a a^b,
\label{TTCC04}
\\
\sbv{\Thetaa}{\Thetaf} 
& = 
- \frac{1}{4} (\rhoa^k_{[a} \partial_k H^{(2)}_{DE|b]}
- C_{ab}^c H^{(2)}_{DEc} ) \eta^A \eta^B a^a a^b
+ \frac{1}{3!} (\rhoa^k_{[a|} \partial_k H^{(1)}_{i|bc]} 
- C_{[ab}^d H^{(1)}_{D|d|c]} ) \eta^D a^a a^b a^c
\nonumber \\ & \quad
- \frac{1}{4!} (\rhoa^k_{[a} \partial_k H^{(0)}_{bcd} 
- C_{[ab}^e H^{(0)}_{e|cd]}) a^a a^b a^c a^d,
\label{TTCC05}
\\
\sbv{\Thetaf}{\Thetaf}  
& = H^{(2)}_{ABa} k^{AC} H^{(2)}_{CDb} \eta^B \eta^D a^a a^b
- \frac{2}{3!} H^{(2)}_{AB[a|} k^{BC} H^{(1)}_{C|bc]} \eta^A a^a a^b a^c
\nonumber \\ & \quad
+ \frac{2}{4!} H^{(1)}_{A[ab|} k^{AB} H^{(1)}_{B|cd]} a^a a^b a^c a^d,
\label{TTCC06}
\end{align}
From Eqs.~\eqref{TTT01}--\eqref{TTT01}, 
and Eqs.~\eqref{TTCC04}--\eqref{TTT06},
we obtain conditions to satisfy $\sbv{\Theta^{\nabla}}{\Theta^{\nabla}}=0$.
For the deformed function 
$\Theta^{\nabla} = \Thetac^{\nabla} + \Thetaa + \Thetaf$,
the homological condition $\sbv{\Theta^{\nabla}}{\Theta^{\nabla}}=0$ 
holds if the following conditions are satisfied,
\begin{align}
& R_{ija}^b =0,
\\
& \rhoe^i_A S_{iab}^c =0,
\\
& \rhoe^j_A \nabla_j \rhoa^i_a - \rhoa^j_a \partial_j \rhoe^i_A -
\rhoe^i_B k^{BC} H^{(2)}_{CAa} =0,
\\
& - \rhoe^i_{[A} \nabla_{[i} H^{(2)}_{BC]a}(x)
+ {}^A \rd_{a} H_{ABC}(x)  = 0,
\\
& - \rhoe^i_{[A} \nabla_{i}H^{(1)}_{B]ab}(x) 
+ {}^A \rd_{[a} H^{(2)}_{AB|b]}(x) 
- H^{(2)}_{[A|C[a|} k^{CD} H^{(2)}_{D|B]|b]} = 0,
\\
& - \rhoe^i_{A} \partial_{i} H^{(0)}_{abc}(x)
+ {}^A \rd_{[a|}H^{(1)}_{A|bc]}(x) 
- H^{(2)}_{AB[a|} k^{BC} H^{(1)}_{C|bc]} = 0,
\\
& - {}^A \rd_{[a}H^{(0)}_{bcd]}(x) 
+ H^{(1)}_{A[ab|} k^{AB} H^{(1)}_{B|cd]}
= 0,
\end{align}
which are a deformation of conditions
\eqref{GCSMLAcondition01}--\eqref{GCSMLAcondition04}.

The AKSZ action \eqref{actionSCSMLA} is modified by
the following terms induced from Eq.~\eqref{fluxterm03},
\begin{align}
S_F
&= \int_{T[1]\Sigma} \rd^3 \sigma \rd^3 \theta
\left[\frac{1}{2} H^{(2)}_{ABa}(\bX) \by^A \by^B \ba ^a 
+ \frac{1}{2} H^{(1)}_{Aab}(\bX) \by^A \ba^a \ba^b
+ \frac{1}{3!} H^{(0)}_{abc}(\bX) \ba^a \ba^b \ba^c
\right].
\end{align}
Thus, the total action is 
\begin{align}
S &= S_C^{\nabla} + S_A + S_F
\nonumber \\ 
&=\int_{T[1]\Sigma} \rd^3 \sigma \rd^3 \theta
\left[- \bz_i^{\nabla} \rd \bX^i 
+ \frac{1}{2} k_{AB} \by^A \rd \by^B
+ \bw_a D \ba^a
\right.
\nonumber \\
& 
\left.
+ \rhoe^i_A \bz_i^{\nabla} \by^A
+ \frac{1}{3!} H_{ABC}(\bX) \by^A \by^B \by^C 
+ \rhoa^i_a(\bX) \bz_i^{\nabla} \ba^a 
- \frac{1}{2} T_{ab}^c(\bX) \ba^a \ba^b \bw_c
\right.
\nonumber \\
& 
\left.
+ \frac{1}{2} H^{(2)}_{ABa}(\bX) \by^A \by^B \ba^a 
+ \frac{1}{2} H^{(1)}_{Aab}(\bX) \by^A \ba^a \ba^b
+ \frac{1}{3!} H^{(0)}_{abc}(\bX) \ba^a \ba^b \ba^c
\right].
\label{GCSMLAfluxaction}
\end{align}

\subsection{General CSMs with boundaries}
\noindent
Suppose that $\Sigma$ has boundaries $\partial \Sigma \neq \emptyset$.
Let $S = S_S^{\nabla} + S_A + S_F$ 
in Eq.~\eqref{GCSMLAfluxaction} be the AKSZ action.
$\sbv{S}{S}$ in the classical master equation does not vanish, and has 
following boundary integrations,
\begin{align}
\sbv{S}{S} &= 2 \int_{T[1]\Sigma} \rd^3 \sigma \rd^3 \theta \
\rd 
\left[- \bz_i^{\nabla} \rd \bX^i 
+ \frac{1}{2} k_{AB} \by^A \rd \by^B
+ \bw_a D \ba^a
\right.
\nonumber \\
& 
\left.
+ \rhoe^i_A \bz_i^{\nabla} \by^A
+ \frac{1}{3!} H_{ABC} \by^A \by^B \by^C 
+ \rhoa^i_a \bz_i^{\nabla} \ba^a 
- \frac{1}{2} T_{ab}^c \ba^a \ba^b \bw_c
\right.
\nonumber \\
& 
\left.
+ \frac{1}{2} H^{(2)}_{ABa} \by^A \by^B \ba^a 
+ \frac{1}{2} H^{(1)}_{Aab} \by^A \ba^a \ba^b
+ \frac{1}{3!} H^{(0)}_{abc} \ba^a \ba^b \ba^c
\right]
\nonumber \\ 
&= 2 \int_{T[1]\partial \Sigma} \rd^2 \sigma \rd^2 \theta
\left[- \bz_i \rd \bX^i 
+ \frac{1}{2} k_{AB} \by^A \rd \by^B
+ \bw_a \rd \ba^a
+ \rhoe^i_A \bz_i \by^A
\right.
\nonumber \\
& 
\left.
+ \rhoe^i_A \omega_{bi}^a \ba^b \bw_a \by^A
+ \frac{1}{3!} H_{ABC} \by^A \by^B \by^C 
+ \rhoa^i_a \bz_i \ba^a 
+ \frac{1}{2} C_{ab}^c \ba^a \ba^b \bw_c
\right.
\nonumber \\
& 
\left.
+ \frac{1}{2} H^{(2)}_{ABa} \by^A \by^B \ba^a 
+ \frac{1}{2} H^{(1)}_{Aab} \by^A \ba^a \ba^b
+ \frac{1}{3!} H^{(0)}_{abc} \ba^a \ba^b \ba^c
\right].
\end{align}
Introduce boundary terms to describe
boundary conditions on the theory.
The following boundary terms are considered here,
\begin{align}
S_b &=\int_{T[1]\partial \Sigma} \rd^2 \sigma \rd^2 \theta
\left[
\frac{1}{2} \mu^{(2)}_{AB}(\bX) \by^A \by^B
+ \mu^{(1)}_{Aa}(\bX) \by^A \ba^a 
+ \frac{1}{2} \mu^{(0)}_{ab}(\bX) \ba^a \ba^b
\right],
\end{align}
where $\mu^{(2)}$ is an $E$ two-form,
$\mu^{(1)}$ is an $E$ one-form taking a values on $A^*$,
$\mu^{(2)}$ is an $E$ zero-form taking a values on $\wedge^2 A^*$.

In order to calculate Poisson brackets including $S_b$, 
we use the following equations for each superfield,
\begin{align}
\delta \bX^i &= \sbv{S}{\bX^i} = \rd \bX^i - \rhoe^i_A \by^A 
- \rhoa^i_a \ba^a,
\label{SPhi04}
\\
\delta \by^A &= \sbv{S}{\by^A} = \rd \by^A
+ k^{AB} \rhoe^i_B \bz_i 
+ k^{AB} \rhoe^i_B \omega_{bi}^a \ba^b \bw_a 
\nonumber \\ & \quad
+ \frac{1}{2} k^{AB} H_{BCD} \by^C \by^D
+ k^{AB} H^{(2)}_{BCa} \by^C \ba^a
+ \frac{1}{2} k^{AB} H^{(1)}_{Bab} \ba^a \ba^b,
\label{SPhi05}
\\
\delta \ba^a &= \sbv{S}{\ba^a} = - \rhoe^i_A \omega^a_{bi} \ba^b \by^A 
+ \rd \ba^a + \frac{1}{2} C^a_{bc} \ba^b \ba^c.
\label{SPhi06}
\end{align}
Using Eqs.~\eqref{SPhi04}--\eqref{SPhi06}, 
$\delta S_b = \sbv{S}{S_b}$ is calculated as
\begin{align}
\delta S_b 
&= \int_{T[1]\partial \Sigma} \rd^2 \sigma \rd^2 \theta
\left[
k^{AC} \rhoe^i_C \mu^{(2)}_{AB} \bz_i^{\nabla} \by^B
+ k^{AB} \rhoe^i_B \mu^{(1)}_{Aa} \bz_i^{\nabla} \ba^a
\right.
\nonumber \\ & \quad
\left.
+ \frac{1}{3!} 
(- \rhoe_{[C|}^i \partial_i \mu^{(2)}_{|AB]} 
+ k^{DE} H_{[AB|D} \mu^{(2)}_{E]C}) \by^A \by^B \by^C
\right.
\nonumber \\ & \quad
\left.
+ \frac{1}{2} (- \rhoa^i_a \partial_i \mu^{(2)}_{AB}
- \rhoe_{[A}^i \nabla_i \mu_{B]a}^{(1)}
+ k^{CD} H_{[A|Ca}^{(2)} \mu^{(2)}_{B]D}
+ k^{CD} H_{[AB|C} \mu^{(1)}_{D]a}) 
\by^A \by^B \ba^a
\right.
\nonumber \\ & \quad
\left.
+ \frac{1}{2} (- \rhoa^i_{[b|} \partial_i \mu^{(1)}_{A|a]}
- C_{ab}^c \mu^{(1)}_{c|a]}
- \rhoe_{A}^i \nabla_i \mu_{ab}^{(0)}
+ k^{BC} H_{Bab}^{(1)} \mu^{(2)}_{CA}
+ k^{BC} H_{AB[a|} \mu^{(1)}_{C|b]}) 
\by^A \ba^a \ba^b
\right.
\nonumber \\ & \quad
\left.
+ \frac{1}{3!} 
(- \rhoa^i_{[a} \partial_i \mu^{(0)}_{bc]}
- C_{[ab|}^d \mu^{(0)}_{d|c]}
+ k^{AB} H_{A[ab|}^{(1)} \mu^{(1)}_{B|c]})
\ba^a \ba^b \ba^c
\right].
\end{align}
$\sbv{S_b}{S_b}$ is
\begin{align}
\sbv{S_b}{S_b}
&= \int_{T[1]\partial \Sigma} \rd^2 \sigma \rd^2 \theta
\left[- k^{AB} \mu^{(2)}_{AB} \mu^{(2)}_{CD} \by^B \by^D
- 2 k^{AC} \mu^{(2)}_{AB} \mu^{(1)}_{Cd} \by^B \ba^d
- k^{AC} \mu^{(1)}_{Aa} \mu^{(1)}_{Cb} \ba^a \ba^b
\right].
\end{align}
We impose boundary conditions
$\bz_i = 0$,
$\bp_i =0$ and $\bw_a =0$ on boundaries.
Then, the condition $\sbv{S_T}{S_T} =0$ for $S_T = S + S_b$ including boundaries 
imposes the following conditions on functions in boundary terms.
\begin{align}
& k^{AC} \rhoe^i_C \mu^{(2)}_{AB} = 0,
\\
& k^{AB} \rhoe^i_B \mu^{(1)}_{Aa} = 0,
\\
& - \rhoe_{[C|}^i \partial_i \mu^{(2)}_{|AB]} 
+ k^{DE} H_{[AB|D} \mu^{(2)}_{E]C} + H_{ABC} = 0,
\\
& - {}^A \rd_a \mu^{(2)}_{AB}
- \rhoe_{[A}^i \nabla_i \mu_{B]a}^{(1)}
+ k^{CD} H_{[A|Ca}^{(2)} \mu^{(2)}_{B]D}
\nonumber \\ &
+ k^{CD} H_{[AB|C} \mu^{(1)}_{D]a}
+ H^{(2)}_{ABa} = 0,
\\
& - {}^A \rd_{[b|} \mu^{(1)}_{A|a]}
- \rhoe_{A}^i \nabla_i \mu_{ab}^{(0)}
+ k^{BC} H_{Bab}^{(1)} \mu^{(2)}_{CA}
\nonumber \\ &
+ k^{BC} H_{AB[a|} \mu^{(1)}_{C|b]}
+ H^{(1)}_{Aab} = 0,
\\
& - {}^A \rd_{[a} \mu^{(0)}_{bc]}
+ k^{AB} H_{A[ab|}^{(1)} \mu^{(1)}_{B|c]}
+ H^{(0)}_{abc} = 0,
\\
& k^{AB} \mu^{(2)}_{AB} \mu^{(2)}_{CD} = 0,
\\
& k^{AC} \mu^{(2)}_{AB} \mu^{(1)}_{Cd} = 0,
\\
& k^{AC} \mu^{(1)}_{Aa} \mu^{(1)}_{Cb]} = 0,
\end{align}

\section{General CSMs with Courant algebroid gauging}
\label{sec:GCSMCAflux}
\noindent
In this section, we consider the general CSM 
with Courant algebroid gauging \eqref{GCSMCAaction},
and introduce flux terms and boundary terms.

\subsection{Introduction of fluxes}
Consider the homological function 
$\Theta^{\nabla} = \Thetac^{\nabla} + \Thetaa$ in 
Eqs.~\eqref{homologicalCA1}--\eqref{homologicalCA2}.
We add following terms to $\Theta^{\nabla}$,
\begin{align}
\Theta_F &=
\frac{1}{2} H^{(2)}_{ABa}(x) \eta^A \eta^B a^a 
+ \frac{1}{2} H^{(1)}_{Aab}(x) \eta^A a^a a^b
+ \frac{1}{3!} H^{(0)}_{abc}(x) a^a a^b a^c.
\label{GCSMCAhomologicalflux}
\end{align}
Here $H^{(2)} = \frac{1}{2} H^{(2)}_{ABa}(x) e^A \wedge e^B \otimes e^a
\in \Gamma(M, \wedge^2 E^* \otimes A^*)$ 
is an $E$-two-form taking a value in $A^*$,
$H^{(1)} = \frac{1}{2} H^{(1)}_{Aab}(x) e^A \otimes e^a \wedge e^b.
\in \Gamma(M, E^* \otimes \wedge^2 A^*)$
is an $E$-one-form taking a value in $\wedge^2 A^*$,
and $H^{(0)}= \frac{1}{3!} H^{(0)}_{abc}(x) e^a \wedge e^b \wedge e^c.
\in C^{\infty}(M, \wedge^3 A^*)$
is an $E$-zero-form taking a value in $\wedge^3 A^*$.

Poisson brackets including $\Thetaf$ are calculated as
\begin{align}
\sbv{\Thetac^{\nabla}}{\Thetaf} 
& = - \frac{1}{3!} \rhoe^i_{[A|} \nabla_{i} H^{(2)}_{|DE]a}
\eta^A \eta^D \eta^E a^a 
+ \frac{1}{4} \rhoe^i_{[A|} \nabla_{i} H^{(1)}_{|D]ab} \eta^A \eta^B a^a a^b
\nonumber \\ & \quad
- \frac{1}{3!} \rhoe^i_{A} \partial_{i} H^{(0)}_{abc} \eta^A a^a a^b a^c
\nonumber \\ & \quad
+ \rhoe^i_A k^{AD}H^{(2)}_{DEa} z^{\nabla}_i \eta^E a^a
+ \frac{1}{2} \rhoe^i_A k^{AD} H^{(1)}_{Dab} z^{\nabla}_i a^a a^b
\nonumber \\ & \quad
+ \frac{1}{3!} H_{[AB|C} k^{CD} H^{(2)}_{D|E]c} \eta^A \eta^B \eta^E a^a
+ \frac{1}{4} H_{ABC} k^{CD} H^{(1)}_{Dab} \eta^A \eta^B a^a a^b,
\label{TTT04}
\\
\sbv{\Thetaa}{\Thetaf} 
& = 
- \frac{1}{4} (\banabla_{[a}  H^{(2)}_{DE|b]}
+ T_{abc} m^{cd} H^{(2)}_{DEd} ) \eta^A \eta^B a^a a^b
\nonumber \\ & \quad
+ \frac{1}{3!} (\banabla_{[a|} H^{(1)}_{D|bc]} 
+ T_{[abd} m^{de} H^{(1)}_{De|c]} ) \eta^D a^a a^b a^c
\nonumber \\ & \quad
- \frac{1}{4!} (\banabla_{[a} H^{(0)}_{bcd} 
+ T_{[abe} m^{ef} H^{(0)}_{f|cd]}) a^a a^b a^c a^d,
\nonumber \\ & \quad
+ \frac{1}{2} \rhoa^i_a m^{ad} H^{(2)}_{DEd} z^{\nabla}_i \eta^D \eta^E
- \rhoa^i_a m^{ad} H^{(1)}_{Dde} z^{\nabla}_i \eta^D a^e
\nonumber \\ & \quad
+ \frac{1}{2} \rhoa^i_a m^{ad} H^{(0)}_{def} z^{\nabla}_i a^e a^f,
\label{TTT05}
\\
\sbv{\Thetaf}{\Thetaf}  
& = H^{(2)}_{ABa} k^{AC} H^{(2)}_{CDb} \eta^B \eta^D a^a a^b
- \frac{2}{3!} H^{(2)}_{AB[a|} k^{BC} H^{(1)}_{C|bc]} \eta^A a^a a^b a^c
\nonumber \\ & \quad
+ \frac{2}{4!} H^{(1)}_{A[ab|} k^{AB} H^{(1)}_{B|cd]} a^a a^b a^c a^d
+ \frac{2}{4!} H^{(2)}_{ABa} m^{ab} H^{(2)}_{CDb} \eta^A \eta^B \eta^C \eta^D
\nonumber \\ & \quad
- \frac{2}{3!} H^{(2)}_{[AB|a} m^{ad} H^{(1)}_{|C]de} \eta^A \eta^B \eta^C a^e
+ \frac{2}{4} H^{(2)}_{ABa} m^{ad} H^{(0)}_{def]} \eta^A \eta^B a^e a^f
\nonumber \\ & \quad
+ \frac{2}{4} H^{(1)}_{Aab} m^{bd} H^{(1)}_{Bde} \eta^A \eta^B a^a a^e
- \frac{2}{3!} H^{(1)}_{A[a|b} m^{bd} H^{(0)}_{d|ef]} \eta^A a^a a^e a^f
\nonumber \\ & \quad
+ \frac{2}{4!} H^{(0)}_{[ab|c} m^{cd} H^{(0)}_{d|ef]} a^a a^b a^e a^f.
\label{TTT06}
\end{align}
We deform the homological function by adding $\Thetaf$,
$\Theta^{\nabla} = \Thetac^{\nabla} + \Thetaa + \Thetaf$.
Combining Eqs.~\eqref{TTT04}--\eqref{TTT06} with 
Eqs.~\eqref{TTT01}--\eqref{TTT03},
$\sbv{\Theta}{\Theta}=0$ is satisfied if the following equations hold,
\begin{align}
& \rhoa^i_a m^{ad} H^{(2)}_{DEd} = 0,
\\
& - \rhoe^j_A \nabla_j \rhoa^i_a + \rhoa^j_a \partial_j \rhoe^i_A 
+ \rhoe^i_C k^{CD}H^{(2)}_{DAa} - \rhoa^i_c m^{cd} H^{(1)}_{Ada} =0,
\\
& \rhoe^i_A k^{AD} H^{(1)}_{Dab} 
+ \rhoa^i_c m^{cd} H^{(0)}_{dab} =0,
\\
& H^{(2)}_{ABa} m^{ab} H^{(2)}_{CDb} =0,
\\
& - {}^A \rd_{a} H_{ABC} 
- \rhoe^i_{[A|} \nabla_{i} H^{(2)}_{|BC]a}
+ H_{[AB|D} k^{DE} H^{(2)}_{E|C]c} 
- H^{(2)}_{[AB|c} m^{cd} H^{(1)}_{|C]da} = 0,
\\
& - 2 \rhoe^i_A \rhoe^j_B R_{ija}^e m_{eb}
+ \rhoe^i_{[A|} \nabla_{i} H^{(1)}_{|D]ab} 
+ H_{ABC} k^{CD} H^{(1)}_{Dab} 
- \rd_{[a} H^{(2)}_{DE|b]}
\nonumber \\ & 
+ H^{(2)}_{C[A|[a|} k^{CD} H^{(2)}_{D|B]|b]} 
+ H^{(2)}_{ABc} m^{cd} H^{(0)}_{dab]}
+ H^{(1)}_{[A|[a|c} m^{dd} H^{(1)}_{|B]d|b]} 
= 0,
\\
& - \rhoe^i_A S_{iabc} - \rhoe^i_{A} \partial_{i} H^{(0)}_{abc} 
+ {}^A \rd_{[a|} H^{(1)}_{A|bc]} 
+ H^{(2)}_{AB[a|} k^{BC} H^{(1)}_{C|bc]} 
- H^{(1)}_{A[a|d} m^{de} H^{(0)}_{e|bc]} 
\\
&
- {}^A \rd_{[a} H^{(0)}_{bcd]} 
+ H^{(1)}_{A[ab|} k^{AB} H^{(1)}_{B|cd]} 
+ H^{(0)}_{[ab|e} m^{ef} H^{(0)}_{f|cd]} =0.
\end{align}
Additional terms in the AKSZ action 
corresponding to Eq.~\eqref{GCSMCAhomologicalflux} are 
\begin{align}
S_F
&= \int_{T[1]\Sigma} \rd^3 \sigma \rd^3 \theta
\left[\frac{1}{2} H^{(2)}_{ABa}(\bX) \by^A \by^B \ba^a 
+ \frac{1}{2} H^{(1)}_{Aab}(\bX) \by^A \ba^a \ba^b
+ \frac{1}{3!} H^{(0)}_{abc}(\bX) \ba^a \ba^b \ba^c
\right].
\end{align}
Then, the total action functional is a deformation 
of the action functional \eqref{GCSMCAaction} with $S_F$,
\begin{align}
S &= S_C^{\nabla} + S_A + S_F
\nonumber \\ 
&=\int_{T[1]\Sigma} \rd^3 \sigma \rd^3 \theta
\left[- \bz_i^{\nabla} \rd \bX^i 
+ \frac{1}{2} k_{AB} \by^A \rd \by^B
+ \frac{1}{2} m_{ab}(\bX) \ba^a D \ba^b
\right.
\nonumber \\
& 
\left.
+ \rhoe^i_A \bz_i^{\nabla} \by^A
+ \frac{1}{3!} H_{ABC}(\bX) \by^A \by^B \by^C
+ \rhoa^i_a(\bX) \bz_i^{\nabla} \ba^a 
+ \frac{1}{3!} T_{abc}(\bX) \ba^a \ba^b \ba^c
\right.
\nonumber \\
& 
\left.
+ \frac{1}{2} H^{(2)}_{ABa}(\bX) \by^A \by^B \ba^a 
+ \frac{1}{2} H^{(1)}_{Aab}(\bX) \by^A \ba^a \ba^b
+ \frac{1}{3!} H^{(0)}_{abc}(\bX) \ba^a \ba^b \ba^c
\right].
\label{GCSMCAactionflux}
\end{align}

\subsection{General CSMs with boundaries}
\noindent
Suppose that $\Sigma$ has boundaries $\partial \Sigma \neq \emptyset$.
Let $S = S_S^{\nabla} + S_A + S_F$ be the AKSZ action in
Eq.~\eqref{GCSMCAactionflux}.
$\sbv{S}{S}$ is not zero because of the following boundary integrations,
\begin{align}
\sbv{S}{S} &= 2 \int_{T[1]\Sigma} \rd^3 \sigma \rd^3 \theta \
\rd 
\left[- \bz_i^{\nabla} \rd \bX^i 
+ \frac{1}{2} k_{AB} \by^A \rd \by^B
+ \frac{1}{2} m_{ab} \ba^a D \ba^b
\right.
\nonumber \\
& 
\left.
+ \rhoe^i_A \bz_i^{\nabla} \by^A
+ \frac{1}{3!} H_{ABC} \by^A \by^B \by^C
+ \rhoa^i_a \bz_i^{\nabla} \ba^a 
+ \frac{1}{3!} T_{abc} \ba^a \ba^b \ba^c
\right.
\nonumber \\
& 
\left.
+ \frac{1}{2} H^{(2)}_{ABa} \by^A \by^B \ba^a 
+ \frac{1}{2} H^{(1)}_{Aab} \by^A \ba^a \ba^b
+ \frac{1}{3!} H^{(0)}_{abc} \ba^a \ba^b \ba^c
\right].
\nonumber \\ 
&= 2 \int_{T[1]\partial \Sigma} \rd^2 \sigma \rd^2 \theta
\left[- \bz_i \rd \bX^i 
+ \frac{1}{2} k_{AB} \by^A \rd \by^B
+ \frac{1}{2} m_{ab} \ba^a \rd \ba^b
\right.
\nonumber \\
& 
\left.
+ \rhoe^i_A \bz_i  \by^A
- \frac{1}{2} \rhoe_A^i \omega_{ai}^c m_{cb} \ba^a \ba^b \by^A
+ \frac{1}{3!} H_{ABC} \by^A \by^B \by^C
\right.
\nonumber \\
& 
\left.
+ \rhoa^i_a \bz_i \ba^a 
+ \frac{1}{3!} C_{abc} \ba^a \ba^b \ba^c
\right.
\nonumber \\
& 
\left.
+ \frac{1}{2} H^{(2)}_{ABa} \by^A \by^B \ba^a 
+ \frac{1}{2} H^{(1)}_{Aab} \by^A \ba^a \ba^b
+ \frac{1}{3!} H^{(0)}_{abc} \ba^a \ba^b \ba^c
\right].
\label{GCSMCAboundary01}
\end{align}
We introduce boundary terms to describe 
consistent boundary conditions on the theory.
We consider following boundary terms,
\begin{align}
S_b &=\int_{T[1]\partial \Sigma} \rd^2 \sigma \rd^2 \theta
\left[
\frac{1}{2} \mu^{(2)}_{AB}(\bX) \by^A \by^B
+ \mu^{(1)}_{Aa}(\bX) \by^A \ba^a 
+ \frac{1}{2} \mu^{(0)}_{ab}(\bX) \ba^a \ba^b
\right].
\end{align}
In order to calculate Poisson brackets including $S_b$, we use the following 
equations for each superfield,
\begin{align}
\delta \bX^i &= \sbv{S}{\bX^i} = \rd \bX^i - \rhoe^i_A \by^A 
- \rhoa^i_a \ba^a,
\label{TTT11}
\\
\delta \by^A &= \sbv{S}{\by^A} = \rd \by^A
+ k^{AB} \rhoe^i_B \bz_i 
- \frac{1}{2} k^{AB} \rhoe^i_B \omega_{ci}^d m_{db} \ba^b \ba^c
\nonumber \\ & \quad
+ \frac{1}{2} k^{AB} H_{BCD} \by^C \by^D
+ k^{AB} H^{(2)}_{BCa} \by^C \ba^a
+ \frac{1}{2} k^{AB} H^{(1)}_{Bab} \ba^a \ba^b,
\label{TTT12} \\
\delta \ba^a &= \sbv{S}{\ba^a} = \rhoe^i_A \omega^a_{bi} \ba^b \by^A 
+ m^{ab} \rhoa_b^i \bz_i + \rd \ba^a + \frac{1}{2} m^{ad} C_{dbc} \ba^b \ba^c.
\nonumber \\ & \quad
+ m^{ab} H^{(2)}_{ABb} \by^A \by^B
+ m^{ab} H^{(1)}_{Acb} \by^A \ba^c
+ \frac{1}{2} m^{ad} H^{(0)}_{dbc} \ba^b \ba^c,
\label{TTT13} 
\end{align}
Using Eqs.~\eqref{TTT11}--\eqref{TTT13}, $\delta S_b = \sbv{S}{S_b}$ is
\begin{align}
\delta S_b 
&= \int_{T[1]\partial \Sigma} \rd^2 \sigma \rd^2 \theta
\left[
k^{AC} \rhoe^i_C \mu^{(2)}_{AB} \bz_i^{\nabla} \by^B
+ k^{AB} \rhoe^i_B \mu^{(1)}_{Aa} \bz_i^{\nabla} \ba^a
\right.
\nonumber \\ & \quad
\left.
- m^{ab} \rhoa^i_b \mu^{(1)}_{Aa} \bz_i^{\nabla} \by^A
+ m^{bc} \rhoe^i_b \mu^{(0)}_{ca} \bz_i^{\nabla} \ba^a
\right.
\nonumber \\ & \quad
\left.
+ \frac{1}{3!} 
(- \rhoe_{[C|}^i \partial_i \mu^{(2)}_{|AB]} 
+ k^{DE} H_{[AB|D} \mu^{(2)}_{E]C}) \by^A \by^B \by^C
\right.
\nonumber \\ & \quad
\left.
+ \frac{1}{2} (- \rhoa^i_a \partial_i \mu^{(2)}_{AB}
- \rhoe_{[A}^i \nabla_i \mu_{B]a}^{(1)}
+ k^{CD} H_{[A|Ca}^{(2)} \mu^{(2)}_{B]D}
\right.
\nonumber \\ & \quad
\left.
+ k^{CD} H_{[AB|C} \mu^{(1)}_{D]a}
+ m^{bc} H_{[A|ab}^{(1)} \mu^{(1)}_{B]c}
+ m^{bc} H_{ABb}^{(2)} \mu^{(0)}_{ca}) 
\by^A \by^B \ba^a
\right.
\nonumber \\ & \quad
\left.
+ \frac{1}{2} (- \rhoa^i_{[b|} \partial_i \mu^{(1)}_{A|a]}
- m^{cd} C_{abc} \mu^{(1)}_{Ad}
+ m^{cd} \rhoa^i_c  \omega_{ai}^e m_{eb} \mu^{(1)}_{Ad}
- \rhoe_{A}^i \nabla_i \mu_{ab}^{(0)}
\right.
\nonumber \\ & \quad
\left.
+ k^{BC} H_{Bab}^{(1)} \mu^{(2)}_{CA}
+ k^{BC} H_{AB[a|} \mu^{(1)}_{C|b]}
+ m^{cd} H_{abc}^{(0)} \mu^{(1)}_{Ad}
+ m^{cd} H_{Ac[a}^{(1)} \mu^{(0)}_{b]d}) 
\by^A \ba^a \ba^b
\right.
\nonumber \\ & \quad
\left.
+ \frac{1}{3!} 
(- \rhoa^i_{[a} \partial_i \mu^{(0)}_{bc]}
+ m^{de} C_{d[ab|} \mu^{(0)}_{e|c]}
+ m^{de} \rhoa^i_d \omega_{bi}^f m_{fc} \mu^{(0)}_{ea}
\right.
\nonumber \\ & \quad
\left.
+ k^{AB} H_{A[ab|}^{(1)} \mu^{(1)}_{B|c]}
+ m^{de} H_{d[ab|}^{(0)} \mu^{(0)}_{e|c]}) 
\ba^a \ba^b \ba^c
\right].
\label{GCSMCAboundary02}
\end{align}
Moreover $\sbv{S_b}{S_b}$ is computed as
\begin{align}
\sbv{S_b}{S_b}
&= \int_{T[1]\partial \Sigma} \rd^2 \sigma \rd^2 \theta
\left[- (k^{CD} \mu^{(2)}_{CA} \mu^{(2)}_{DB} 
+ m^{ab} \mu^{(1)}_{Aa} \mu^{(1)}_{Bb}) \by^A \by^B
\right.
\nonumber \\ & \quad
\left.
- 2 (k^{BD} \mu^{(2)}_{BA} \mu^{(1)}_{Dc} 
- m^{ab} \mu^{(1)}_{Aa} \mu^{(0)}_{bc}) \by^A \ba^c
- ( k^{AB} \mu^{(1)}_{Aa} \mu^{(1)}_{Bb} 
+ m^{cd} \mu^{(0)}_{ac} \mu^{(0)}_{bd]}) \ba^a \ba^b
\right].
\label{GCSMCAboundary03}
\end{align}
We impose boundary conditions, $\bz_i = 0$,
and half of degrees of freedom of both $\by^A$ and $\ba^a$ are zero.
Concretely, two inner products vanish $k_{AB} \by^A \rd \by^B 
= m_{ab} \ba^a \rd \ba^b =0$ on boundaries.

Using Eqs.~\eqref{GCSMCAboundary01}--\eqref{GCSMCAboundary03},
the condition $\sbv{S_T}{S_T}=0$ 
imposes the following conditions to $\mu^{(0)}$,
$\mu^{(1)}$ and $\mu^{(2)}$ in boundary terms,
\begin{align}
& k^{AC} \rhoe^i_C \mu^{(2)}_{AB} = 0,
\\
& k^{AB} \rhoe^i_B \mu^{(1)}_{Aa} = 0,
\\
& m^{ab} \rhoa^i_b \mu^{(1)}_{Aa} = 0,
\\
& m^{bc} \rhoe^i_b \mu^{(0)}_{ca} = 0,
\\
& - \rhoe_{[C|}^i \partial_i \mu^{(2)}_{|AB]} 
+ k^{DE} H_{[AB|D} \mu^{(2)}_{E]C} + H_{ABC} = 0,
\\
& - {}^A \rd_a \mu^{(2)}_{AB}
- \rhoe_{[A}^i \nabla_i \mu_{B]a}^{(1)}
+ k^{CD} H_{[A|Ca}^{(2)} \mu^{(2)}_{B]D}
\nonumber \\ & 
+ k^{CD} H_{[AB|C} \mu^{(1)}_{D]a}
+ m^{bc} H_{[A|ab}^{(1)} \mu^{(1)}_{B]c}
+ m^{bc} H_{ABb}^{(2)} \mu^{(0)}_{ca}
+ H^{(2)}_{ABa} = 0,
\\
& 
- {}^A \rd^{\nabla}_{[b|} \mu^{(1)}_{A|a]}
- \rhoe_{A}^i \nabla_i \mu_{ab}^{(0)}
+ k^{BC} H_{Bab}^{(1)} \mu^{(2)}_{CA}
+ k^{BC} H_{AB[a|} \mu^{(1)}_{C|b]}
\nonumber \\ & 
+ m^{cd} H_{abc}^{(0)} \mu^{(1)}_{Ad}
+ m^{bc} H_{ABb}^{(2)} \mu^{(0)}_{ca}
+ H^{(1)}_{Aab} = 0,
\\
& - {}^A \rd^{\nabla}_{[a} \mu^{(0)}_{bc]}
+ k^{AB} H_{A[ab|}^{(1)} \mu^{(1)}_{B|c]}
+ m^{de} H_{d[ab|}^{(0)} \mu^{(0)}_{e|c]}
+ H^{(0)}_{abc} = 0,
\\
& k^{CD} \mu^{(2)}_{AC} \mu^{(2)}_{BD} + 
m^{ab} \mu^{(1)}_{Aa} \mu^{(1)}_{Bb} =0,
\\
& k^{BD} \mu^{(2)}_{BA} \mu^{(1)}_{Dc} 
- m^{ab} \mu^{(1)}_{Aa} \mu^{(0)}_{bc} =0,
\\
& k^{AB} \mu^{(1)}_{Aa} \mu^{(1)}_{Bb} 
+ m^{cd} \mu^{(0)}_{ac} \mu^{(0)}_{bd} =0.
\end{align}

\newcommand{\bibit}{\sl}




\end{document}